\documentclass[aps,amsmath,amssymb,nofootinbib,longbibliography]{revtex4-2} 
\usepackage{graphicx} 
\usepackage{amsmath}
\usepackage{amsfonts,amsbsy}
\usepackage{amssymb}
\usepackage{relsize}
\allowdisplaybreaks

\usepackage{xcolor}
\definecolor{lcolor}{rgb}{0.5,0,0}
\definecolor{citcolor}{rgb}{0,0.3,0.0}
\usepackage[breaklinks,colorlinks,urlcolor=blue,citecolor=citcolor,linkcolor=lcolor]{hyperref}

\usepackage{mciteplus}

\def\be{\begin{equation}}
\def\ee{\end{equation}}
\def\bea{\begin{eqnarray}}
\def\eea{\end{eqnarray}}

\newcommand{\dd}{{\rm d}}
\newcommand{\nn}{\nonumber}
\newcommand{\tr}{\mathrm{tr} }

\begin{document}

\title{
  Entanglement entropy of the proton in coordinate space}

\author{Adrian Dumitru}
\affiliation{Department of Natural Sciences, Baruch College, CUNY,
17 Lexington Avenue, New York, NY 10010, USA}
\affiliation{The Graduate School and University Center, The City University
  of New York, 365 Fifth Avenue, New York, NY 10016, USA}
\author{Alex Kovner}
\affiliation{Physics Department, University of Connecticut, 2152 Hillside Road, Storrs, CT 06269, USA}
\author{Vladimir V. Skokov}
\affiliation{North Carolina State University, Raleigh, NC 27695, USA}
\affiliation{RIKEN-BNL Research Center, Brookhaven National Laboratory, Upton, NY 11973, USA}

\begin{abstract}
  We calculate the entanglement entropy of a model proton wave
  function in coordinate space by integrating out degrees of freedom
  outside a small circular region $\bar A$ of radius $L$, where $L$ is
  much smaller than the size of the proton. The wave function provides
  a nonperturbative distribution of three valence quarks. In addition, we
  include the perturbative emission of a single gluon and calculate
  the entanglement entropy of gluons in $\bar A$. For both, quarks and
  gluons we obtain the same simple result: $S_E =-\int\frac{dx}{\Delta
    x}\, N_{L^2}(x)\log[N_{a^2}(x)]$, where $a$ is the UV cutoff in
  coordinate space and $\Delta x$ is the longitudinal resolution
  scale. Here $N_{S}(x)$ is the number of partons (of the appropriate
  species) with longitudinal momentum fraction $x$ inside an area
  $S$. It is related to the standard parton distribution function
  (PDF) by $N_S(x)=\frac{S}{A_p}\, \Delta x\, F(x)$, where $A_p$ denotes
  the transverse area of the proton.
\end{abstract}

\maketitle
\tableofcontents

\section{Introduction}

Rapid advent of quantum science in recent years provides strong
motivation for asking new types of questions in many areas of inquiry,
including high energy nuclear and particle physics. In particular,
there is an ongoing vigorous discussion about the relevance of
entanglement (and the associated entanglement entropy) in the context
of particle production in high energy hadronic
collisions~\cite{Kovner:2015hga, Kovner:2018rbf, Armesto:2019mna,
  Duan:2020jkz, Duan:2021clk, Duan:2023zls, Hagiwara:2017uaz, Kharzeev:2017qzs,
  Kharzeev:2021yyf, Kharzeev:2021nzh, Dvali:2021ooc, Tu:2019ouv,Ramos:2020kaj, Hentschinski:2021aux, Hentschinski:2022rsa, Muller:2022htn, Ehlers:2022oal, Kou:2022dkw, Ramos:2022gia, Liu:2022hto, Liu:2022bru, Dumitru:2022tud, Dumitru:2023fih}.

The initial discussion by Kharzeev and Levin~\cite{Kharzeev:2017qzs}
is framed in the context of entanglement of the degrees of freedom
inside a small area of the proton actually probed in a DIS experiment,
with the rest of the degrees of freedom in the proton wave function,
and, in particular, with soft modes of the gluon field responsible for
confinement. It was suggested that the entropy of this entanglement
translates into the Boltzmann entropy of particles produced in the
collision. Some model calculations have been performed to probe this
picture~\cite{Tu:2019ouv, Ramos:2020kaj, Kharzeev:2021yyf,
  Hentschinski:2021aux, Hentschinski:2022rsa}, and it has also been
 subjected to an experimental test~\cite{H1:2020zpd}. However, no
direct calculation of entanglement entropy in coordinate space has so
far been reported in the literature. The aim of this manuscript is to fill
this gap.

Of course, such a calculation requires knowledge of the wave function
of the proton, and needless to say, the exact proton wave function is
not known. Nevertheless, several simple model wave functions that
provide the distribution of valence quarks at large $x$ and low
resolution $Q^2$ have been used in QCD phenomenology over the years
with reasonable success,
e.g.\ refs.~\cite{Schlumpf:1992vq,Brodsky:1994fz,Xu:2021wwj,
  Shuryak:2022thi}. These quark wave functions can be improved by
including a perturbative gluon component, as described in
ref.~\cite{Dumitru:2020gla}, and used in ref.~\cite{Dumitru:2023sjd}
to compute DIS structure functions at high energy, and in
ref.~\cite{Dumitru:2022tud} to study entanglement of momentum-space
degrees of freedom over the whole area of the proton. In this paper
our main goal is to derive expressions for the density matrix and
entropy of a small ``hole'' in the proton in such a setup. For
numerical estimates we will use one specific light-cone valence quark
model wave function from refs.~\cite{Schlumpf:1992vq,Brodsky:1994fz}.
 
The idea of our calculation is very straightforward. We divide the
transverse area of the proton into a small disc $\bar A$ and its
complement $A$, and integrate out all degrees of freedom in $A$.  The
result is the reduced density matrix $\rho_{\bar A}$ which contains
complete information for the calculation of any observable localized
in $\bar A$. We then calculate the von Neumann entropy of
$\rho_{\bar A}$.

Even before including the perturbative gluon component, the result is
nontrivial. The entropy in this case is associated with different
numbers of quarks that can reside inside $\bar A$. Note that the total
number of valence quarks in the model wave function is fixed (three),
nevertheless the wave function carries finite probabilities of finding
different numbers of quarks inside $\bar A$. Integrating over $A$
therefore generates a reduced density matrix which spans Hilbert
subspaces with different occupation numbers, $n$. The von~Neumann
entropy arises precisely due to nonvanshiing eigenvalues of
$\rho_{\bar A}$ in subspaces with different $n$.

Note that in the simple case when the total number of partons is
fixed, the reduced density matrix is diagonal in the particle number
basis {\it by fiat}. This follows immediately since in reducing the
density matrix we trace over $A$, and thus calculate matrix elements
between states which have equal numbers of partons in $A$. For wave
functions that do not preserve the number of partons we expect, in
general, that $\rho_{\bar A}$ would not be diagonal in the $n$
basis. Thus, including a perturbative gluon emission may lead to such a
nondiagonal $\rho_{\bar A}$. As it turns out, in the first order of
perturbation theory this does not happen due to the fact that in the
valence part of the wave function the color and spatial degrees of
freedom are not entangled with each other.

We first perform the calculation in the way described above for the
valence wave function that contains three quarks only. We next
include a one gluon state which is generated by the first order
perturbative correction. Here, for simplicity we modify our procedure
somewhat, i.e.\ we trace over the quark degrees of freedom in the
whole wave function, and only then do we generate $\rho_{\bar A}$ by
reducing over the gluon degrees of freedom in $A$. We then calculate
the entanglement entropy of the resulting density matrix, which now has the meaning of entropy of gluons inside $\bar A$.

This paper is structures as follows. In Section~\ref{sec:groundwork}
we prepare our tools for performing the calculation in coordinate
space and describe the model wave function for valence quarks. In
Section~\ref{sec:rho_Abar-qqq} we calculate the reduced density matrix
$\rho_{\bar A}$ and the entropy for a small disc $\bar A$ in this
model. Here ``small" means small relative to the nonperturbative scale
which determines the spatial size of the model wave function. We
discuss the dependence of the entropy on the area of $\bar A$ in this
regime. In Section~\ref{sec:n-partDM} we include an additional
perturbatively emitted gluon in the wave function, and again calculate
the reduced density matrix (in the way described above) and discuss
its properties. The entanglement entropy is also calculated in
sec.~\ref{sec:EEpert-rho}.  In both cases (quarks and gluons) the
entanglement entropy can be written in a very suggestive form in terms
of the PDF of the appropriate parton species, eq.~\eqref{sb}.
Finally, in Section~\ref{sec:Discussion} we discuss our results and
their possible relation to the suggestion of
ref.~\cite{Kharzeev:2017qzs}.

\section{Laying the groundwork}  \label{sec:groundwork}

In the following we denote any three vector $p$ as $p=(p^+,\vec p)$,
where $p^+$ and $\vec p$ are longitudinal and transverse components of
the vector respectively.  We will be using a mixed representation for
the wave function where coordinate space is used to represent the
transverse degrees of freedom, and momentum space for the longitudinal
ones.

In this mixed representation we denote a state of the proton at center
of mass (COM) position $\vec R=0$ and longitudinal momentum $P^+$ by $|\vec
R=0, P^+\rangle$. This convoluted notation does not reference the wave
function for the internal degrees of freedom, i.e.\ the coordinates, color
and spin states of the constituents, which we will specify in a short
while.

The coordinate space proton state vector is related to the momentum space
state vector through (see e.g.\ ref.~\cite{Burkardt:2002hr})
\be
|\vec R, P^+\rangle = {\cal N} \int_{\vec P} \, e^{i \vec P \cdot \vec R}\,
|\vec P, P^+\rangle ~,
\ee
where $\vec P$ is the transverse momentum of the proton, and the
integration measure is
\be
\int_{\vec P} \equiv \int \frac{\dd^2 P}{(2\pi)^2}~.
\ee
The normalization factor is determined from the condition  $|{\cal N}|^2\, \int_{\vec P}=1$.
A proton centered at $\vec R=0$ is then
\be \label{eq:p-centered-R=0}
|\vec R=0, P^+\rangle = {\cal N} \int_{\vec P} \, |\vec P, P^+\rangle ~.
\ee
We employ the standard normalization of the momentum space states:
\be \label{eq:<p|k>_norm}
\left< K \, | \, P \right> = 16\pi^3 \, P^+\, \delta(P^+ - K^+)
\, \delta^2(\vec P - \vec K)~
\ee
which leads to the following normalization of the 
the mixed space state vector 
\be  \label{eq:norm_p-R}
\langle \vec R=0, P^{\prime +}\, |\, \vec R=0, P^+\rangle =
4\pi \, P^+\, \delta(P^+ - P^{\prime +})~.
\ee
The density operator for this state is
\be
\hat \rho = |\vec R=0, P^+\rangle\,\, \langle \vec R=0, P^+ |~.
\ee
In the following we will be calculating matrix elements of $\hat\rho$
between states of the partonic (Fock) Hilbert space
\be \label{eq:rho_alpha_alpha'}
\rho_{\alpha\alpha'} = 
\langle \alpha'|\vec R=0, P^+\rangle
\,\,
\langle \vec R=0, P^+ | \alpha\rangle~,
\ee
where $\alpha$ denotes a collection of ``labels'' (such as the LC
momentum fractions $x_i$, coordinates and color indices) assigned to
the basis vectors of the Fock space.

\subsection{The valence quark Fock state}
\label{sec:qqqDM-LO}

We start with considering states that contain three valence quarks
only. In the model described below the color and spatial degrees of
freedom are not entangled, i.e.\ the wave function is a direct product
of the color and spatial state vectors. In this case, for the spatial
wave function $\alpha=\{x_i,\vec r_i\}$ refers to the quark LC
momentum fractions and their transverse coordinates.

The state vector $|P^+,\vec P\rangle$ of a proton made of $N_c$
``valence'' quarks is written as
\be \label{Pstate}
|P\rangle = \sum_{h_i}\, \int\limits_{[0,1]^{N_c}}
 [\dd x_i] \int [\dd^2 k_i]\,
\Psi\left(k_i, h_i\right)\,\,
\frac{1}{\sqrt{N_c!}}\sum_{i_1\dots i_{N_c}} \epsilon_{i_1\cdots i_{N_c}}
\left|p_1, i_1, h_1; \cdots; p_{N_c},i_{N_c},h_{N_c}\right>~,
\ee
where
\bea
\left[\dd x_i\right] &=& \delta\left(1-\sum_i x_i\right) \,\,
  \prod_{i}\frac{\dd x_i}{2x_i}~,  \label{eq:[dx_i]}\\
  \left[\dd^2 k_i\right] &=& 
  (2\pi)^3\, \delta\left(\sum_i \vec k_i\right) \,\,
  \prod_{i}\frac{\dd^2 k_i}{(2\pi)^3}~.   \label{eq:[dk_i]}
\eea
Here $k_i=(k_i^+,\vec{k}_i)$ denote the momenta of the $i$-th quark in the
transverse rest frame of the proton, and $\vec p_i = \vec k_i +
x_i\vec P$. The space-helicity wave function $\Psi\left(k_i,
h_i\right)$ is symmetric under exchange of any two quarks while the
state is antisymmetric in color space. In what follows we will mainly
focus on the spatial wave function and trace out spin-flavor and
color degrees of freedom.

We can now write the proton state in terms of the quark Fock space states
\be \label{eq:|R=0>_|p1p2p3>}
|\vec R=0, P^+\rangle = {\cal N} \int_{\vec P} \, 
\int [\dd x_i] \int [\dd^2 k_i]\,
\Psi\left(k_i\right)\,\,
\left|p_1 ; \, p_2; \, p_3\right>~,
\ee
where we have omitted the quark (and proton) spins, for simplicity. 
Summing up, we integrate over the Galilean-invariant ``internal'' quark
transverse momenta subject to the constraint that they add up to zero,
and then over the COM transverse momentum $\vec P$, which is also the momentum of the proton.\\

Analogously, the three-quark coordinate space state with the quarks
located at $\vec r_i$ and carrying LC momentum fractions $x_i$ is
constructed as:
\be \label{eq:|r>_|q>}
|x_1, \vec r_1;\,x_2, \vec r_2;\,x_3, \vec r_3 \rangle =
{\cal N} \int_Q \int [\dd^2 q_i] \, e^{-i \sum (\vec q_i+x_i\vec Q)\cdot \vec r_i}\,
|x_i, \vec q_i+x_i\vec Q\rangle~.
\ee
%
Equation~(\ref{eq:|r>_|q>}) can be extended to four (and more) particles
simply by adding labels for momentum fraction and transverse
position/momentum of the additional particle to the state vector, and
including the momentum of the additional particle in the integration
measure eq.~(\ref{eq:[dk_i]}).

The overlap of the proton state with the state of three quarks
localized at fixed transverse coordinates is given by
\bea
\langle \vec R=0, P^+| x_i, \vec r_i\rangle &=&
|{\cal N}|^2\, \int_{P,Q} \int [\dd y_i] \int [\dd^2 k_i]\,
\int [\dd^2 q_i]\, e^{-i\sum (\vec q_i+x_i\vec Q)\cdot \vec r_i}
\, \Psi^*(y_i,\vec k_i) \\
& & ~~~~~
\prod_i
\langle y_i, \vec k_i+y_i\vec P \, |\, x_i, \vec q_i+x_i\vec Q \rangle \nonumber\\
&=& |{\cal N}|^2\,
(2\pi)^3\, \delta(1-\sum x_i)\,\, \delta(\sum x_i\vec r_i)\,
\int [\dd^2 q_i] \, e^{-i \sum \vec q_i\cdot \vec r_i}\,
\Psi^*(x_i, \vec q_i)~,\nonumber
\label{eq:<R|r>}
\eea
where we used eqs.~(\ref{eq:<p|k>_norm},\ref{eq:[dx_i]},
\ref{eq:[dk_i]}).  Note that the overlap does not vanish only for
states with COM located at the origin, $\sum x_i \vec r_i=0$, just as
for the proton, c.f.\ eq.~(11) in~\cite{Burkardt:2002hr}; or
ref.~\cite{Hatta:2017cte} for the analogous case of a $q\bar q$ dipole. Also, the LC
momentum fractions of the quarks must sum up to one.  Since only such
states contribute to the proton density matrix, and we included the
constraints on the longitudinal momentum fractions/the transverse
momenta in the integration measure (\ref{eq:[dx_i]},\ref{eq:[dk_i]}),
a matrix element of the properly normalized density matrix is given by
\bea
\rho_{\alpha\alpha'} &=&
\frac{\langle \vec R=0, P^+| \alpha'\rangle}
{|{\cal N}|^2\, (2\pi)^3\, \delta(1-\sum x'_i)\,\, \delta(\sum x'_i\vec r_i\!')}
\,
\frac{\langle \alpha | \vec R=0, P^+\rangle}
{|{\cal N}|^2\, (2\pi)^3\, \delta(1-\sum x_i)\,\, \delta(\sum x_i\vec r_i)}
     \label{eq:rho_aa'}
\\
&=&
\int[\dd^2 q_i] \, e^{i \sum \vec q_i\cdot \vec r_i}\,
\int[\dd^2 q_i'] \, e^{-i \sum \vec q_i'\cdot \vec r_i\!'}\,
\Psi^*(x_i', \vec q_i')\,\,
\Psi(x_i, \vec q_i)   \label{eq:rho_r_r'}\\
&=& \Psi^*(x_i', \vec r_i\!')\,\, \Psi(x_i, \vec r_i)~,
\label{eq:rho_r_r'_coord}
\eea
where
$\alpha=\{x_i, \vec r_i|\, \sum x_i=1, \sum x_i\vec r_i=0\}$ and
$\alpha'=\{x_i', \vec r_i\!'|\, \sum x_i'=1, \sum x_i'\vec r_i\!'=0\}$
denote two sets of LC momentum fractions and transverse quark
positions.
Here
in the last step we used the definition (B.4) of
ref.~\cite{Burkardt:2002hr} for the coordinate space LC wave
functions:
\be
\Psi(x_i, \vec r_i) = \int[\dd^2 q_i] \, e^{i \sum \vec q_i\cdot \vec r_i}\,
\Psi(x_i, \vec q_i)~.
\label{eq:Psi_position_space}
\ee
The normalization of the coordinate space wave function will be
obtained later in eq.~(\ref{eq:norm_tr-Abar}) from the requirement
that the trace of the density matrix $\tr\, \hat \rho=1$.

For the model wave function considered here (see below)
the color degrees of freedom of the above density matrix could be
restored simply by
multiplying by the normalized color space matrix
$
\frac{1}{3!}\, \epsilon_{i_1 i_2 i_3}\, \epsilon_{i_1' i_2' i_3'}$.
%

\subsection{A model wave function}  \label{sec:BSwf}

Our main goal here is to obtain general expressions for the reduced
density matrix in a transverse region $\overline A$ of the
proton and to estimate the entropy associated with this density matrix (which we do in  sec.~\ref{sec:EE_qqq}). For this we require an explicit expression for the
three-quark wave function $\Psi_\mathrm{qqq}$.

We employ a simple model due to Schlumpf and
Brodsky~\cite{Schlumpf:1992vq,Brodsky:1994fz},
\begin{equation}
  \Psi\left(x_i,\vec k_i\right)
  \sim
  \, \sqrt{x_1 x_2 x_3}\,\,
  e^{-{\cal M}^2/2\beta^2}; \ \ \ \ \ \ {\cal M}^2 = \sum \frac{\vec k_i^2+m_q^2}{x_i}\,.
  \label{eq:Psiqqq_HO}
\end{equation}
Here ${\cal M}^2 $ is the invariant mass squared of the
non-interacting three-quark system~\cite{Bakker:1979eg}, i.e.\ the sum
of the quark LC energies multiplied by $P^+$. The non-perturbative
parameters $m_q=0.26$~GeV and $\beta=0.55$~GeV have been fixed in
Refs.~\cite{Schlumpf:1992vq,Brodsky:1994fz} to match empirical
properties of the proton at low energy and low resolution.
Note that $\beta$ is of order $N_c=3$ times the root-mean-square
valence quark transverse momentum in the proton.\\

This  Gaussian wave function can be easily transformed to
position space. One obtains (up to normalization)
\be \label{eq:Psi_qqq-coord-space}
\Psi\left(x_i,\vec r_i\right) \sim
F(x_1,x_2,x_3)\,\, e^{-\frac{1}{2} \, a_{13}\, \beta^2\, r_{13}^2} \,\,
e^{-\frac{1}{2} \, a_{23}\, \beta^2 \, r_{23}^2}
\,\, e^{b\, \beta^2\, \vec r_{13}\cdot \vec r_{23}}
\ee
with
\bea
\vec r_{ij} &\equiv & \vec r_i - \vec r_j~, \nn\\
F(x_1,x_2,x_3) &=& (2\pi\beta^2)^2\, \frac{(x_1 x_2 x_3)^{3/2}}{(2\pi)^6}\,
\,\,
e^{-\frac{m_q^2}{2\beta^2}\sum \frac{1}{x_i}}~, \nn\\
a_{23} &=& x_2\, (1-x_2)~,\nn\\
a_{13} &=& x_1\, (1-x_1)~,\nn\\
b &=& x_1\, x_2~.
\eea
One can easily verify that this is symmetric under the exchange of any
two quarks, $(x_i,\vec r_i) \leftrightarrow (x_j,\vec r_j)$; $i,
j=1,2,3$.

\section{The reduced density matrix and entanglement entropy of a three quark system}  \label{sec:rho_Abar-qqq}

We can now construct a reduced density matrix by tracing over a
subset of degrees of freedom.  Here we are interested in the reduced
density matrix that determines observables localized to a small
circle in the center of the proton. To find this density matrix we
have to trace over the region $A$ of the proton which is the outside
of the circle in question. In other words we have to integrate over
the transverse positions and LC momentum fractions of all quarks
located in $A$.\\

\subsection{The density matrix for a small disc}
First we note that the Hilbert space inside the disc $\bar A$ is a
direct sum of Hilbert spaces of zero, one, two and three particles. In
addition, it is obvious that since we are tracing over $A$, the reduced
density matrix does not contain off diagonal elements that connect
states with different particle numbers.  The reduced density matrix
therefore can be represented as a block diagonal matrix of the form
\be
 \rho_{\overline A} \,=\, 
\begin{pmatrix}
    \rho_0 & 0 & 0 & 0\\
0 &   \rho_1 & 0 & 0 \\
0 & 0 &   \rho_2 & 0 \\
0 & 0 & 0 &   \rho_3 \\
\end{pmatrix}
\label{eq:rho_Abar}
\ee
Note that the various blocks along the diagonal are density matrices
over Hilbert spaces of different dimensionality.

To calculate $\rho_0$ we place all quarks in $A$,
\be \label{eq:rho_0}
\rho_0 = \int[\dd x_i]
\int [\dd^2 r_i]\,\, \Theta_A(\vec r_1)\,\Theta_A(\vec r_2)\,
\Theta_A(\vec r_3)\,\,|\Psi(x_i,\vec r_i)|^2~.
\ee
Here,
\be
[\dd^2 r_i]=\dd^2 r_1\,\dd^2 r_2\,\dd^2 r_3\,\delta(\sum x_i\vec r_i)~,
\ee
and $\Theta_A(\vec r)=1$ if $\vec r\in A$ and 0
otherwise.  This is a pure dimensionless (by the normalization
condition in eq.~(\ref{eq:norm_tr-Abar}) below) number giving the
probability that in our wave function no quarks reside in $\bar A$.

The second block $\rho_1$ of~(\ref{eq:rho_Abar}) is the probability
density that only one of the quarks is localized in $\overline A$
while the other two are localized in $A$.  Tracing over $A$ we have to
set $\vec r_1=\vec r_1\!'\in A$ and $\vec r_2=\vec r_2\!'\in A$, so by
virtue of the COM constraint we also have $\vec r_3=\vec r_3\!'$, with
$\vec r_3\in \overline A$, so $\rho_1$ is diagonal in coordinate
space:
\be
(\rho_1)_{\alpha\alpha} = 3
\int \frac{\dd x_1\dd x_2}{8 x_1 x_2 x_3}\,
\delta\left(1-\sum x_i\right)
\int \dd^2r_1\, \dd^2r_2\, \delta\left(\sum
x_i\vec r_i\right)\,\Theta_A(\vec r_1)\,\Theta_A(\vec r_2)\,\, |\Psi(x_i,\vec r_i)|^2
~,
~~~~~~~(\vec r_3\in \overline A).
\label{eq:rho11}
\ee
The matrix indices here are $\alpha=\{x_3, \vec r_3\}$, defined over
the domain $0\le x_3\le1$ and $\vec r_3\in \overline A$.

Clearly, the dimensionalities of $\rho_1$ and $\rho_0$ are
different. While $\rho_0$ is dimensionless and has the meaning of
probability, $\rho_1$ has dimension $1/r^2$ and has the meaning of
probability density. To construct a probability from $\rho_1$ we would
have to multiply it by the ``lattice spacing" in the transverse
coordinate space, $a^2$, and in fact also by the elementary length in
the longitudinal momentum space, $\Delta x$. If we take this route, the
integration over the coordinate $\vec r_3$ and the momentum fraction
$x_3$ will have to be performed with the dimensionless measure
$\dd^2r_3/a^2\ \dd x_3/\Delta x$.

For the discussion of the density matrix itself this is not
crucial since a calculation of the average of any observable involves
integration over $x$ and $r_i$ and the minimal area cancels in the
product of the probability density and the integration
measure. However when we calculate von~Neumann entropies
$S_E$ this becomes important, since we need to define a dimensionless
{\it probability} in order to take its logarithm. In fact it is also
crucial to work with a dimensionless density matrix when we calculate
the trace of any nontrivial (not first) power of $\rho$.  Since the index
on $\rho_1$ is continuous, the density matrix is infinitely
dimensional. We therefore expect its individual matrix elements to
vanish in the strict continuum limit (for vanishing $a^2$ and $\Delta
x$) as the first power of $a^2\Delta x$. When calculating
$\tr\, \rho_1$ this smallness of the matrix elements is compensated by
the integration over $\vec r_3, \ x_3$. However when we calculate $\tr\,
\rho_1^N$, the diagonal matrix elements now vanish as $(a^2\Delta
x)^N$, while there is still only a single integral over $\vec r_3,\ x_3$
involved in calculating the trace. Therefore $\tr\,
\rho_1^N\rightarrow_{a,\Delta x\rightarrow 0} (a^2\Delta
x)^{N-1}$, and it is imperative to keep the lattice spacing
finite in order to obtain any physical information about $\tr\,
\rho_1^N$ beyond the trivial fact that it vanishes in the continuum
limit. We will therefore introduce the lattice spacing in the
definition of the density matrix and will forthwith work with
\be
(\rho_1)_{\alpha\alpha} = 3\Delta x \, a^2
\int \frac{\dd x_1\dd x_2}{8 x_1 x_2 x_3}\,
\delta\left(1-\sum x_i\right)
\int \dd^2r_1\, \dd^2r_2\, \delta\left(\sum
x_i\vec r_i\right)\,\Theta_A(\vec r_1)\,\Theta_A(\vec r_2)\,\, |\Psi(x_i,\vec r_i)|^2
~,
~~~~~~~(\vec r_3\in \overline A)~,
\label{eq:rho1}
\ee
with the understanding that the trace is taken with respect to the
measure $\frac{\dd^2r_3}{a^2}\frac{\dd x_3}{\Delta x}
\Theta_{\overline A}(\vec r_3)$.  Furthermore, we  included the
$x_3$-dependent part of the integration measure~(\ref{eq:[dx_i]}),
i.e.\ the factor $1/x_3$, in the definition of $\rho_1$ so that the
trace is given by the $x_3$-independent integration measure $\frac{\dd
  x_3}{\Delta x}$. One can easily understand why this is necessary by
considering the classical Shannon entropy of a propability density
distribution, see\ appendix \ref{sec:app_Shannon}.\\

The third block $\rho_2$ corresponds to the configuration where two of the
quarks are located in $\overline A$ while the third one is located in $A$:
\be \label{eq:rho_2}
(\rho_2)_{\alpha\alpha'} = 3\,(a^2\Delta x)^2\,
\frac{\Psi^*(\alpha',x_3,\vec r_3)}{x_3\sqrt{8 x_1' x_2' x_3}}\,
\frac{\Psi(\alpha,x_3,\vec r_3)}{x_3\sqrt{8 x_1 x_2 x_3}}
~,
\ee
with $\alpha=\{x_1,\vec r_1, \, x_2, \vec r_2\}$ and $\alpha' =
\{x_1',\vec r_1\!', \, x_2', \vec r_2\!'\}$.  Note that there is no
integral over the coordinate of the third quark in this equation. This
is due to the fact that COM constraint rigidly determines $\vec
r_3, x_3$ for given coordinates and longitudinal momenta of the first
two quarks as
\be
x_3=1-x_1-x_2=1-x_1'-x_2', \ \ \ \ \ \ \ \vec r_3=-(x_1\vec r_1+x_2\vec
r_2)/x_3=-(x_1\vec r_1\!' +x_2\vec r_2\!')/x_3~.
\ee
The matrix indices $\alpha, \alpha'$ are defined over the domain where
these relations are satisfied with $0\le x_3\le1$ and $\vec r_1, \vec r_2,
\vec r_1\!', \vec r_2\!' \in \overline A$, $\vec r_3\in A$.  We have again
introduced the lattice spacing into the definition of a matrix element
of $\rho_2$ to make it dimensionless.  The factor $3$ in~(\ref{eq:rho_2})
arises since either one of the three quarks can reside in $A$.

The trace of $\rho_2$ on the subspace with two particles is defined as
\bea
\tr\, \rho_2 &=&
\int\frac{\dd x_1}{\Delta x}\frac{\dd x_2}{\Delta x}\,
\int \frac{\dd^2r_1}{a^2}\frac{\dd^2r_2}{a^2}\,
\Theta_{\overline A}(\vec r_1)\,\Theta_{\overline A}(\vec r_2)\,
\Theta(x_3)\, \Theta_A(\vec r_3)\,
(\rho_2)_{\alpha\alpha} \nn\\
&=& 3 \int [\dd x_i]\, \int [\dd^2r_i]\,
\Theta_A(\vec r_3)\,\Theta_{\overline A}(\vec r_1)\,\Theta_{\overline A}(\vec r_2)\,
\, |\Psi(x_i,\vec r_i)|^2~,
\eea
where the lattice spacing cancels between the matrix element and the
integration measure. Once again, the trace operation does not involve
any Jacobians which depend on $x_1$ and $x_2$.\\

Finally, the fourth block corresponds to all three quarks in $\overline A$:
\be
(\rho_3)_{\alpha \alpha'} =
\frac{(a^2\, \Delta x)^2}{x_3'\sqrt{8 x_1' x_2' x_3'}}\,
\frac{1}{x_3\sqrt{8 x_1 x_2 x_3}}\,\,
\Psi^*(x_i',\vec r_i\!')\, \Psi(x_i,\vec r_i)~.
\ee
Here $\alpha=\{x_1,\vec r_1, x_2,\vec r_2\}$, and similarly for
$\alpha'$.  These indices are defined over the domain $0 \le x_1, x_2,
x_1', x_2' \le 1$ with $0\le x_3=1-x_1-x_2\le 1$, $0\le
x_3'=1-x_1'-x_2'\le 1$; and $\vec r_1,\vec r_1\!',\vec r_2,\vec
r_2',\vec r_3,\vec r_3\!' \in \overline A$, with $\vec r_3 = -(x_1\vec
r_1+x_2\vec r_2) /x_3$, $\vec r_3\!' = -(x_1'\vec r_1\!' +x_2'\vec r_2\!')
/x_3'$.  We have again introduced the lattice spacing in this
definition so that the elements of $\rho_3$ are dimensionless,
although these factors cancel in the trace of an arbitrary power of
$\rho_3$.  To take the trace in this block we calculate
\bea
\tr\, \rho_3 &=&
\int \frac{\dd x_1}{\Delta x}\frac{\dd x_2}{\Delta x}
\int \frac{\dd^2 r_1}{a^2}\frac{\dd r_2}{a^2}\,
\Theta(x_3)\, \Theta_{\overline A}(\vec r_1)\,\Theta_{\overline A}(\vec r_2)\,
\Theta_{\overline A}(\vec r_3)\, (\rho_3)_{\alpha \alpha}
\nn\\
&=&
\int [\dd x_i]\, [\dd^2r_i]\,
\left|\Psi(x_i,\vec r_i)\right|^2\, \prod\Theta_{\overline A}(\vec r_i)~.
\eea
Putting this all together we obtain that the total trace of the
density matrix is\footnote{Note that on account of the permutation
  symmetry of the wave function, the following equality holds when
  multiplied by $|\Psi(\vec r_1,\vec r_2,\vec r_3)|^2$ under the
  integral : $\Theta_A(\vec r_1)\,\Theta_A(\vec r_2)\,\Theta_A(\vec
  r_3) + 3 \Theta_A(\vec r_1)\,\Theta_A(\vec r_2)\,\Theta_{\overline
    A}(\vec r_3) + 3 \Theta_A(\vec r_1)\,\Theta_{\overline A}(\vec
  r_2)\,\Theta_{\overline A}(\vec r_3) + \Theta_{\overline A}(\vec
  r_1)\,\Theta_{\overline A}(\vec r_2)\,\Theta_{\overline A}(\vec r_3)
  = 1$. }
\be
\tr\, \rho_{\overline A} = \rho_0 + \tr\,\rho_1 +\tr\, \rho_2
+\tr\, \rho_3 = \int[\dd x_i]\,[\dd^2r_i]\,\, |\Psi(x_i,\vec r_i)|^2 = 1\,.
\label{eq:norm_tr-Abar}
\ee
The normalization of the coordinate space wave function is
determined from this relation.  \\~~\\

In Appendix B we present expressions for calculating traces of powers
of $\rho$ which illustrate explicitly the need to introduce the
``lattice spacing" in our calculation.

\subsection{Entanglement entropy}  \label{sec:EE_qqq}

We now discuss the von~Neumann entropy associated with tracing the
pure state $|\vec R=0, P^+\rangle\, \langle\vec R=0, P^+|$ over the
area $A$:
\be
S_\mathrm{vN} = - \lim_{\epsilon\to0} \frac{\tr\,( \rho_{\bar A})^{1+\epsilon}-1}{\epsilon}
~.
\ee
Because we performed a partial trace over a {\em pure state}, this
entropy represents a measure for the entanglement of the degrees of
freedom remaining in $\overline A$ with those from region $A$, which
have been traced out. We discuss the nature of entanglement in more
detail in the following sec.~\ref{sec:what_is_entangled}.

Using the expressions from Appendix B for $N=1+\epsilon$ and expanding
to linear order in $\epsilon$ this gives
\bea
- S_\mathrm{vN} &=& \rho_0\log\rho_0 + \tr\,\rho_3\log\tr\,\rho_3 \nn\\
  & &+ 3\int[\dd x_i]\,[\dd^2 r_i]\, \Theta_{\overline A}(\vec r_3)\,
\Theta_{A}(\vec r_1)\,\Theta_{A}(\vec r_2)\,
|\Psi(x_i,\vec r_i)|^2 \nn\\
& & \hspace{2cm}\times\log\left( 3\Delta x\, a^2
\int[\dd y_i]\,[\dd^2 s_i]\, \delta(\vec s_3-\vec r_3)\,
\delta(x_3-y_3)\,
\Theta_{A}(\vec s_1)\,\Theta_{A}(\vec s_2)\,
|\Psi(y_i,\vec s_i)|^2\right) \nn\\
& &+ 3 \int[\dd x_i]\,[\dd^2 r_i]\, \Theta_{A}(\vec r_3)\,
\Theta_{\overline A}(\vec r_1)\,\Theta_{\overline A}(\vec r_2)\,
|\Psi(x_i,\vec r_i)|^2 \nn\\
& & \hspace{2cm}\times \log\left( 3\Delta x\, a^2
\int[\dd y_i]\,[\dd^2 s_i]\, \delta(\vec s_3-\vec r_3)\,
\delta(x_3-y_3)\,
\Theta_{\overline A}(\vec s_1)\,\Theta_{\overline A}(\vec s_2)\,
|\Psi(y_i,\vec s_i)|^2\right)
 ~,    \label{eq:SvN}
\eea
where we used  $\tr\,\rho_{\overline A}=1$.

This is a rather formal expression, and to understand some of its
properties we will consider the dependence of the entropy on the area
$\overline A$ of the cutout.

When the region $\overline A$ shrinks to a point we of course expect
the entropy to vanish\footnote{The same is true if $\overline A$ encompasses
the entire transverse space.}. Indeed, all the terms in eq.~(\ref{eq:SvN})
vanish, except $\rho_0$ which approaches 1: for vanishingly small area
the probability to find zero particles inside is unity. Taking the
area small but nonvanishing, for a circular cutout with radius $L$, we
have:
\bea \label{eq:dS0/dL}
\frac{\partial S_\mathrm{vN}^{(0)}}{\partial L} &=&
- \frac{\partial\rho_0}{\partial L} = 2\pi L\, \int\dd x\, I(x)
~~~~~~~~~~~~~~~~(\mathrm{for}\, L\to0) \\
I(x) &=& 3 \int[\dd y_i]\,\delta(y_3-x)
\int \dd^2r_1\, \dd^2r_2\,\, \delta(x_1\vec r_1+x_2\vec r_2)\,\,
|\Psi(y_1,\vec r_1; y_2, \vec r_2; y_3,\vec 0)|^2~.
\label{eq:I(x)}
\eea
For the model wave function from sec.~\ref{sec:BSwf} we obtain the
numerical estimate
\be \label{eq:dS0/dL_coefficient}
\frac{1}{\beta}\, \frac{\partial S_\mathrm{vN}^{(0)}}{\partial L}
= 2\pi \,L\beta\, \cdot 0.534(1)~.
\ee
\\

There is an additional contribution of order $\sim L^2$ to the
entropy. It is due to the third term in eq.~(\ref{eq:SvN}) which
originates from $\rho_1$. Again, for a circular cutout $\overline A$
of radius $L$ centered at the origin, we have
\bea\label{s1q}
\frac{\partial S_\mathrm{vN}^{(1)}}{\partial L} &=&
2\pi L\, \int\dd x\, I(x)\,
\log\frac{1}{a^2\,(\Delta x)\, I(x)}~,~~~~~~~(\mathrm{for}\, L\to0)~.
\eea
For small ``lattice spacing'' $a$ the logarithm in this expression is
large, and this is in fact the dominant contribution to the derivative
of the entropy with respect to $L$. For example, for the model wave
function from sec.~\ref{sec:BSwf}, and for a fairly coarse resolution
of transverse position and longitudinal momentum, $\Delta x\,
(a\beta)^2=0.1$, we obtain the numerical estimate
\be  \label{eq:dS1/dL_coefficient}
\frac{1}{\beta}\, \frac{\partial S_\mathrm{vN}^{(1)}}{\partial L}
= 2\pi \,L\beta\, \cdot 1.21(1) ~.
\ee
Thus, for small $L^2$ the leading contribution to the entropy is
\bea\label{s1qq}
S_\mathrm{vN}^{(1)} &=&-
\pi L^2\, \int\dd x\, I(x)\,
\log [a^2\,(\Delta x) I(x)]~.
\eea
This can be rewritten in a more transparent way if we notice that
$I(x)$ as defined in \eqref{eq:I(x)} is nothing but the density of
quarks with longitudinal momentum $x$ in the proton $I(x)=F(x)/A_p$,
where $A_p$ is the transverse area of the proton and $F(x)$ is the quark
PDF. We then have
\bea\label{sFqq}
S_\mathrm{vN}^{(1)} &=&-
\frac{\pi L^2}{A_p}\, \int\dd x\, F(x)\,
\log [\frac{a^2}{A_p}\,(\Delta x) F(x)]\,.
\eea
The dependence of the entropy on the lattice spacing is easily
understood. Since $\rho_1$ is a matrix with continuous index, we
expect its eigenvalues to be small, i.e.\ of order $a^2$, while the
number of nonvanishing eigenvalues is large $O(1/a^2)$. For such a
matrix with a large number of small eigenvalues, the entropy is indeed
proportional to the logarithm of the inverse eigenvalue, and this is
what we see in \eqref{s1q}. The area scaling of the entropy is also
quite natural, since at small $L$ the number of degrees of freedom in
the reduced density matrix is proportional to the area of the cutout.

\subsection{What is entangled here?}
\label{sec:what_is_entangled}

We would now like to comment on the nature of entanglement that
produces the entanglement entropy that we calculated. It is
somewhat different from the naive picture of entanglement we are used
to in a vacuum state of a Quantum Field Theory. In the QFT setting one
divides space into two regions $A$ and $\bar A$ and considers
the wave function of local field degrees of freedom in the two regions
$\Phi(x\in \bar A)$ and $\Phi(y\in A)$. The entanglement is then
understood in terms of nonfactorizability of the wave function
$\Psi[\Phi(x),\Phi(y)]\ne \Psi_1[\Phi(x)]\Psi_2[\Phi(y)]$, and the
entanglement entropy is associated with this nonfactorizability.

In our case the nature of entanglement is somewhat different. It is
not that some internal degree of freedom of quarks in $A$, like color
or helicity, is entangled with quarks in $\bar A$. In fact, we do not
have to consider several quarks with internal degrees of freedom at
all to understand our result. Let us imagine having just one quark in
the proton area. This quark can be either in $A$ or in $\bar A$. We
can write the total wave function of the quark in terms of the basis
states in the Hilbert spaces $H_A$ and $H_{\bar A}$. For simplicity, we
will even forget about different transverse coordinates in $A$ and
$\bar A$. The wave function of our quark can then be written as
\begin{equation}
\Psi=a|0\rangle_A\times|1\rangle_{\bar A}+b|1\rangle_A\times|0\rangle_{\bar A}
\end{equation}
where $|a|^2$ is the probability that the quark is in $\bar A$ and
$|b|^2=1-|a|^2$ is the probability that it is in $A$. Tracing over $A$
removes the relative phase of $a$ and $b$ and we generate the reduced
density operator $\hat \rho_{\bar A}=\Big[|a|^2|1\rangle\langle
  1|+|b|^2|0\rangle\langle 0|\Big]_{\bar A}$. This is a mixed state
over $\bar A$ and carries the entanglement entropy. Thus, the entanglement in our calculation is between the
quark being (or not being) in $A$ and the {\bf same} quark being (or
not being) in $\bar A$. These states are maximally entangled since the
total number of quarks is fixed to be exactly one. This is a ``quantum
mechanical" rather than ``QFT type'' entanglement, very similar to the
``Schr\"odinger cat" thought
experiment~\cite{Schrodinger:1935q,Schrodinger:1936q}, where one
should read {\bf one quark in $A$} as {\it the cat is alive }, and
{\bf no quark in $\bar A$} as {\it radioactive nucleus intact}; {\bf
  no quark in $A$} as {\it the cat is dead} and {\bf the quark is in
  $\bar A$} as {\it radioactive nucleus decayed}.

\section{Including the $|qqqg\rangle$ Fock state}
\label{sec:n-partDM}

We now add the $|qqqg\rangle$ Fock states into our calculation. In
perturbation theory,  such states have nonvanishing
probability at order $g^2$. We write the proton state schematically in
the form
\be\label{pure}
|P\rangle \sim
\Psi_{qqq}\, \epsilon_{i_1 i_2 i_3}\, |q_{i_1}\,q_{i_2}\,q_{i_3}\rangle
+
\Psi_{qqqg}\, \left[(t^a)_{j i_1} \epsilon_{i_1 i_2 i_3}\,
|q_{j}\,q_{i_2}\,q_{i_3}\, g_a\rangle-(i_1\leftrightarrow i_2)-(i_1\leftrightarrow i_3)\right]~.
\ee
In the leading perturbative order the three quark wave function
$\Psi_{qqq}$ includes the $O(g^2)$ virtual corrections, and
$\Psi_{qqqg}$ is the (3 quarks+1 gluon) spatial wave function at order
$O(g)$.

In the two components the quarks are in different representations of
color-SU(3): they are in the color singlet in the
$|qqq\rangle$ state, and in the color octet in $|qqqg\rangle$.

In the following we will calculate the entanglement entropy for the
same geometry as in the previous section. To simplify the calculations,
however, we will trace the density matrix over the colors of the quarks.
The pure state \eqref{pure} is described by a density operator which
in principle contains off diagonal matrix elements in the particle
number basis
\bea
|P\rangle \, \langle P| &\sim&
\Psi_{qqq}\, \epsilon_{i_1 i_2 i_3}\, \Psi^*_{qqq}\, \epsilon_{i  i_2' i_3'}\,
|q_{i_1}\,q_{i_2}\,q_{i_3}\rangle\, \langle q_{i_1'}\,q_{i_2'}\,q_{i_3'}| \nn\\
& & +
\Psi_{qqqg}\, (t^a)_{j i_1} \epsilon_{i_1 i_2 i_3}\,
\Psi^*_{qqqg}\, (t^{a^\prime})_{i_1' j'} \epsilon_{i_1' i_2' i_3'}\,
|q_{j}\,q_{i_2}\,q_{i_3}\, g_a\rangle\,
\langle q_{j'}\,q_{i_2'}\,q_{i_3'}\, g_{a'}| \nn\\
& & +
\Psi_{qqq}\, \epsilon_{i_1 i_2 i_3}\,
\Psi^*_{qqqg}\, (t^{a^\prime})_{i_1' j'} \epsilon_{i_1' i_2' i_3'}\,
|q_{i_1}\,q_{i_2}\,q_{i_3}\rangle\, \langle q_{j'}\,q_{i_2'}\,q_{i_3'}\, g_{a'}|
\nn\\
& & +
\Psi_{qqqg}\, (t^a)_{j i_1} \epsilon_{i_1 i_2 i_3}\,
\Psi^*_{qqq}\, \epsilon_{i_1' i_2' i_3'}\,
|q_{j}\,q_{i_2}\,q_{i_3}\, g_a\rangle\,
\langle q_{i_1'}\,q_{i_2'}\,q_{i_3'}|
~.
\eea
However, the off diagonal matrix elements vanish after tracing over
quark colors precisely due to the fact that the three quarks are in
the color singlet state in $\Psi_{qqq}$ and in the color octet state
in $\Psi_{qqqg}$.  The reduced (over the quark color) density operator
is diagonal in particle number and has the form
\bea
\tr_{qqq-\mathrm{colors}} \hat\rho &\sim&
3!\, \Psi_{qqq}\, \Psi^*_{qqq}\,
|q\,q\,q\rangle\, \langle q\,q\,q| \nn\\
& & + \delta^{a{a^\prime}}\,
\Psi_{qqqg}\, \Psi^*_{qqqg}\,
|q\,q\,q\, g\rangle\,
\langle q\,q\,q\, g|~.
\label{eq:tr_q-colors_rhoNLO}
\eea
%

\subsection{$\Psi_{qqqg}$ at order $g$ and $\Psi_{qqq}$ at order $g^2$}

Our first order of business is to calculate the perturbative wave
function. For simplicity, we restrict ourselves to the soft gluon
approximation, i.e.\ we assume that the gluon longitudinal momentum is
much smaller than the typical  longitudinal momentum of a quark.\\

Let us begin with $\Psi_{qqqg}$. The emission of a gluon from one of
the quarks generates the following ${\cal O}(g)$
correction\footnote{We are following the notation and expressions from
  refs.~\cite{Dumitru:2020gla,Dumitru:2022tud}.} to the momentum space
proton state $|P\rangle$:
\bea \label{Pstate_O(g)}
|P^+,\vec P\rangle_{{\cal O}(g)} &=& g \int [\dd x_i] \int [\dd^2 k_i]\,
\Psi_\mathrm{qqq}(k_i)\,\,
\frac{1}{\sqrt 6} \sum_{j_1 j_2 j_3} \epsilon_{j_1 j_2 j_3}
\int\limits_{\Delta x} \frac{\dd x_g}{2x_g}\frac{\dd^2 k_g}{(2\pi)^3}
\sum_{\sigma m a} \nn\\
& &
\left[(t^a)_{m j_1}  \frac{\Theta(x_1-x_g)}{x_1-x_g} \,
\hat\psi_{q\to qg}(p_1; p_1-k_g, k_g)
|m,p_1-k_g;\, j_2,p_2;\, j_3,p_3\rangle \right.\nn\\
& & +
(t^a)_{m j_2}  \frac{\Theta(x_2-x_g)}{x_2-x_g} \,
\hat\psi_{q\to qg}(p_2; p_2-k_g, k_g)
|j_1,p_1;\, m,p_2-k_g;\, j_3,p_3\rangle \nn\\
& & \left. +
(t^a)_{m j_3}  \frac{\Theta(x_3-x_g)}{x_3-x_g} \,
\hat\psi_{q\to qg}(p_3; p_3-k_g, k_g)
|j_1,p_1;\, j_2,p_2;\, m,p_3-k_g\rangle\right]\otimes |a,k_g, x_g, \sigma\rangle~.
\eea
The integration measures here, $[\dd x_i]$ and $[\dd^2 k_i]$, pertain to
coordinates of the {\em parent} quarks. We have cut off the integration
over the light-cone momentum fraction of the gluon $x_g$ by $\Delta x$
to regularize the soft singularity in QCD. That is, we prohibit
gluon emission into the lowest ``bin'' of $x_g$.

The light-cone gauge Fock space amplitude for the $qg$ state of a
quark in the soft gluon approximation in $D=4$ dimensions is
\be
\hat\psi_{q\to qg}(p; k_q, k_g) =
2\, \frac{x_p}{k_g^2 + \Delta^2}\,
\vec k_g \cdot \vec \epsilon^*_\sigma
\ee
where $x_p=p^+/P^+$, and $\Delta^2$ is a regulator for the collinear
singularity. Physically, the regularization is provided by the finite
size of the color singlet state which the emitter is a part of. Thus
the magnitude of the regulator $\Delta$ is of order $\Lambda_{\rm QCD}$, or in
our case of the order of the inverse size of the model proton wave
function set by the parameter $\beta$. It is much smaller that the
inverse radius squared of the cutout $\bar A$.

Projecting on the Fock space state $|\alpha\rangle$, where $\alpha$
denotes a set of four momentum fractions $x_i$, transverse positions
$\vec r_i$ and colors $i_1, i_2, i_3, a$, we obtain
\bea
\langle\alpha|P^+,\vec R=0\rangle_{{\cal O}(g)} &=& 2g\,
\frac{|{\cal N}|^2}{(2\pi)^2}   \,
(2\pi)^3\, \delta\left(1-\sum x_i\right)\, \delta\left(\sum x_i \vec r_i\right)
\frac{1}{\sqrt 6} \int [\dd^2 k_i]\, e^{i\sum \vec k_i\cdot\vec r_i}
\frac{\vec k_g\cdot\vec \epsilon_\sigma^*}{k_g^2+\Delta^2}
\nn\\
& &
\sum_{j}\left[
  \epsilon_{j i_2 i_3}\, (t^a)_{i_1 j}\,\,
\Psi_\mathrm{qqq}(k_1+k_g; k_2; k_3)
+ \epsilon_{i_1 j i_3}\, (t^a)_{i_2 j}\,\,
\Psi_\mathrm{qqq}(k_1;k_2+k_g; k_3)
\right. \nn\\
& & ~~~~~~~~~~~~ \left. + \epsilon_{i_1 i_2 j}\, (t^a)_{i_3 j}\,\,
\Psi_\mathrm{qqq}(k_1; k_2; k_3+k_g)
\right]~.
\label{eq:<alpha|R=0>_g}
\eea

To properly account for probability conservation we also need to
include the ${\cal O}(g^2)$ virtual corrections to
$\Psi_\mathrm{qqq}$. There are two types of such corrections.  The
first one arises due to emission and reabsorption of a gluon by one of
the quarks, and amounts to multiplying the momentum space quark state
vectors in eq.~(\ref{eq:|R=0>_|p1p2p3>}) by the wave function
renormalization factor
\be
\left(Z_q(x_1)\,Z_q(x_2)\,Z_q(x_3) \right)^{1/2} =
1 - \frac{1}{2}\left(C_q(x_1) + C_q(x_2) + C_q(x_3)\right)~,
\ee
with
\bea
C_q(x_1) &=&
\frac{g^2C_F}{4\pi^2} \int_{\Delta x/x_1}^{1}
\frac{\dd z}{z}\, A_0(\Delta^2)~,
\label{eq:Cq}\\
A_0(\Delta^2) &=& 4\pi \int \frac{\dd^2n}{(2\pi)^2}
\frac{1}{\vec{n}^2 + \Delta^2}~.  \label{eq:A0_div}\nonumber
\eea
Again, a cutoff $\Delta x$ on the momentum fraction of the gluon was introduced here.  We regulate $A_0(\Delta^2)$ in the UV by a
Pauli-Villars type regulator
\bea
A_0^{\mathrm{reg}}(\Lambda^2/\Delta^2) &=& A_0(\Delta^2) - A_0(\Lambda^2) =
4\pi \int \frac{\dd^2n}{(2\pi)^2}
\left[\frac{1}{\vec{n}^2 + \Delta^2} - \frac{1}{\vec{n}^2 + \Lambda^2} \right]
= \log\frac{\Lambda^2}{\Delta^2}~,
\label{eq:A0_reg}
\eea
where $\Lambda^2$ is a UV cutoff. Then,
\be \label{eq:C_q-reg}
C^{\mathrm{reg}}_q(x_1) = \frac{g^2C_F}{4\pi^2} \, \log\frac{x_1}{\Delta x}\,
\log\frac{\Lambda^2}{\Delta^2}~.
\ee
\\
We were forced to introduce the momentum UV regulator in the
present calculation in order to regulate gluon emissions at short
transverse distances. Recall that earlier we had to introduce a
similar (coordinate space) regulator $a$ in order to define
probabilities and entropy for a continuous system, e.g.\ in
\eqref{eq:rho1}. The two regulators of course should not be considered
independent. In the following we take them to be related as
$\Lambda^2=1/a^2$, in the same way as we took the regulator of the
soft divergence to be equal to the ``lattice spacing" in the
longitudinal momentum space $\Delta x$.

The second virtual correction to $\Psi_\mathrm{qqq}$ is due to the exchange
of a gluon between any pair of  quarks.  Let quark 1 emit and
quark 2 absorb the gluon in $|P\rangle$; we then have (again for $x_g\to0$):
\bea
|P^+,\vec P\rangle^{12}_{{\cal O}(g^2)} &=& \int [\dd x_i] \int [\dd^2 k_i]\,
\Psi_\mathrm{qqq}\left(k_1; k_2; k_3\right)\,\,
\frac{1}{\sqrt 6}\sum_{j_1 j_2 j_3} \epsilon_{j_1 j_2 j_3}\nn\\
& &
g^2\sum_{\sigma, a, n, m} (t^a)_{m j_1} (t^a)_{n j_2}
\int_{\Delta x} \frac{\dd x_g}{2x_g}
\frac{\dd^2 k_g}{(2\pi)^3} \frac{1}{x_1} \,
\hat\psi_{q\to qg}(p_1; p_1-k_g, k_g) \nn\\
& &
\frac{1}{x_2} \,
\hat\psi_{qg\to q}(p_2, k_g; p_2+k_g)
\,
\left|m,p_1-k_g;\, n,p_2+k_g;\, j_3,p_3\right>
~.      \label{eq:Pstate_O(g2Xchange)}
\eea
Here, the amplitude for the absorption of a gluon by a quark is
\be
\hat\psi_{qg\to q}(k_q, k_g; p) = -
2x_p\, \frac{\vec k_g \cdot \vec \epsilon_\sigma}{k_g^2+\Delta^2}~.
\ee
We can now sum over gluon polarizations, $\sum_\sigma
\vec k_g \cdot \vec \epsilon_\sigma\!^*\, \vec k_g \cdot \vec \epsilon_\sigma =
k_g^2$. Changing variables, $\vec k_1\to \vec k_1 + \vec k_g$ and
$\vec k_2\to \vec k_2 - \vec k_g$,
we obtain
\bea
|P^+,\vec P\rangle^{12}_{{\cal O}(g^2)} &=& - 4g^2 \int [\dd x_i] \int [\dd^2 k_i]\,
\frac{1}{\sqrt 6}\sum_{j_1 j_2 j_3} \epsilon_{j_1 j_2 j_3}
\sum_{a, n, m} (t^a)_{m j_1} (t^a)_{n j_2} \nn\\
& &
\int_x \frac{\dd x_g}{2x_g}
\frac{\dd^2 k_g}{(2\pi)^3}
\Psi_\mathrm{qqq}\left(k_1+k_g; k_2-k_g; k_3\right)\,\,
\frac{k_g^2}{(k_g^2 + \Delta^2)^2}
\,
\left|m,p_1;\, n,p_2;\, j_3,p_3\right>
~.      \label{eq:Pstate_O(g2Xchange)b}
\eea

Adding analogous contributions corresponding to gluon exchanges
between quarks 1, 3, and 2, 3 we finally have
\bea
|P^+,\vec R=0\rangle_{{\cal O}(g^2)} &=& - 4g^2 \int [\dd x_i]
\int [\dd^2 r_i]\,
\frac{1}{\sqrt 6}\sum_{j_1 j_2 j_3} \epsilon_{j_1 j_2 j_3}
\int_{\Delta x} \frac{\dd x_g}{2x_g}
\frac{\dd^2 k_g}{(2\pi)^3}\, \frac{k_g^2}{(k_g^2 + \Delta^2)^2}
\int [\dd^2 k_i]\, e^{i\sum\vec k_i\cdot \vec r_i}\,\,
\sum_{a, n, m} \nn\\
& & \left[ (t^a)_{m j_1} (t^a)_{n j_2}
\Psi_\mathrm{qqq}\left(k_1+k_g; k_2-k_g; k_3\right)\,\,
\left|m,x_1,\vec r_1;\, n,x_2,\vec r_2;\, j_3,x_3,\vec r_3\right> \right.\nn\\
& & + (t^a)_{m j_1} (t^a)_{n j_3}
\Psi_\mathrm{qqq}\left(k_1+k_g; k_2; k_3-k_g\right)\,\,
\left|m,x_1,\vec r_1;\, j_2,x_2,\vec r_2;\, n,x_3,\vec r_3\right> \nn\\
& & \left. + (t^a)_{m j_2} (t^a)_{n j_3}
\Psi_\mathrm{qqq}\left(k_1; k_2+k_g; k_3-k_g\right)\,\,
\left|j_1,x_1,\vec r_1;\, m,x_2,\vec r_2;\, n,x_3,\vec r_3\right> \right]
~.      \label{eq:Pstate_O(g2Xchange)_R}
\eea
Projecting this onto the three quark Fock state $\langle j_i, x_i,
\vec r_i|$ gives
\bea
\langle j_i, x_i, \vec r_i|P^+,\vec R=0\rangle_{{\cal O}(g^2)} &=&
-4g^2\, \frac{|{\cal N}|^2}{(2\pi)^2}\,
\delta\left(1-\sum x_i\right)\, (2\pi)^3\, \delta\left(\sum
x_i \vec r_i\right)\,\int[\dd^2 q_i]\, e^{i\sum\vec q_i\cdot\vec r_i}\,
\int_{\Delta x} \frac{\dd x_g}{2x_g}\frac{\dd^2 k_g}{(2\pi)^3}\,
\frac{k_g^2}{(k_g^2+\Delta^2)^2}\nn\\
& & \sum_{i_1 i_2 i_3}
\left[\frac{1}{\sqrt 6}\epsilon_{i_1 i_2 j_3}
(t^a)_{j_1 i_1}(t^a)_{j_2 i_2}
\Psi_\mathrm{qqq}\left(x_1,\vec q_1+\vec k_g; x_2,\vec q_2-\vec k_g;
x_3, \vec q_3\right) \right.\nn\\
& &
+\frac{1}{\sqrt 6}\epsilon_{i_1 j_2 i_3}
(t^a)_{j_1 i_1}(t^a)_{j_3 i_3}
\Psi_\mathrm{qqq}\left(x_1,\vec q_1+\vec k_g; x_2,\vec q_2;
x_3, \vec q_3-\vec k_g\right) \nn\\
& &
\left. + \frac{1}{\sqrt 6}\epsilon_{j_1 i_2 i_3}
(t^a)_{j_2 i_2}(t^a)_{j_3 i_3}
\Psi_\mathrm{qqq}\left(x_1,\vec q_1; x_2,\vec q_2+\vec k_g;
x_3, \vec q_3-\vec k_g\right) \right]
~.   \label{eq:<x_i,r_i|R=0>_g2}
\eea
Alternatively, in terms of the position space wave function,
\bea
\langle j_i, x_i, \vec r_i|P^+,\vec R=0\rangle_{{\cal O}(g^2)} &=&
-4g^2\, \frac{|{\cal N}|^2}{(2\pi)^2}\,
\delta\left(1-\sum x_i\right)\, (2\pi)^3\, \delta\left(\sum
x_i \vec r_i\right)\,
\Psi_\mathrm{qqq}\left(x_i,\vec r_i\right)
\int_{\Delta x} \frac{\dd x_g}{2x_g}\frac{\dd^2 k_g}{(2\pi)^3}\,
\frac{k_g^2}{(k_g^2+\Delta^2)^2}\nn\\
& & \sum_{i_1 i_2 i_3}
\left[\frac{1}{\sqrt 6}\epsilon_{i_1 i_2 j_3}
(t^a)_{j_1 i_1}(t^a)_{j_2 i_2}\, e^{-i\vec k_g\cdot(\vec r_1-\vec r_2)}\,
 \right.\nn\\
& &
+\frac{1}{\sqrt 6}\epsilon_{i_1 j_2 i_3}
(t^a)_{j_1 i_1}(t^a)_{j_3 i_3}\, e^{-i\vec k_g\cdot(\vec r_1-\vec r_3)}\,
 \nn\\
& &
\left. + \frac{1}{\sqrt 6}\epsilon_{j_1 i_2 i_3}
(t^a)_{j_2 i_2}(t^a)_{j_3 i_3}\, e^{-i\vec k_g\cdot(\vec r_2-\vec r_3)}\,
 \right]
~.   \label{eq:<x_i,r_i|R=0>_g2_r-space}
\eea
%

\subsection{First perturbative correction to the density matrix}

We are now in a position to calculate the perturbative correction to
the density matrix.

As already mentioned at the beginning of this section,
eq.~(\ref{eq:tr_q-colors_rhoNLO}), after tracing over the colors of
the quarks the density matrix takes the form
\be
\rho \,=\, 
\begin{pmatrix}
\rho^{qqq} & 0 \\
0 & \rho^{qqqg} \\
\end{pmatrix}\,.
\label{eq:rhoNLO-traced}
\ee
Note that since $\rho^{qqq}$ and $\rho^{qqqg}$ are probability
densities on subspaces with different numbers of particles, they have
different dimensions.  The trace operations over the two entries are
given by $\dd x_1/(2x_1)\,\dd x_2/(2x_2)\,\dd
x_3/(2x_3)\,\delta\left(1-\sum_i x_i\right)\, \dd^2 r_1\,\dd^2 r_2\,\dd^2
r_3\,\delta(\sum x_i\vec r_i)$ and $\dd x_1/(2x_1)\,\dd
x_2/(2x_2)\,\dd x_3/(2x_3)\,\delta\left(1-\sum_i x_i\right)\, \dd x_g/(2x_g)\,
\dd^2 r_1\,\dd^2 r_2\,\dd^2 r_3\,2\pi\dd^2 r_g\, \delta(\sum x_i\vec
r_i)$, for $\rho^{qqq}$ and $ \rho^{qqqg}$, respectively. Hence, if one
is interested in probabilities given by $\rho$ (or its purity, entropy
etc.) one must multiply $\rho^{qqqg}$ by $2\pi \, a^2\Delta x/2x_g$,
as we did in the previous sections.

Let us now compute the matrix~(\ref{eq:rhoNLO-traced}). We begin with
$\rho^{qqq}$ which gives the probability density on the three-quark
state Hilbert space and includes ${\cal O}(g^2)$ virtual
corrections. The first correction is to multiply the ${\cal O}(1)$
non-perturbative density matrix from eq.~(\ref{eq:rho_aa'}) by six
wave function renormalization factors, $\prod Z^{1/2}(x_i)=1 - \sum
C^{\mathrm{reg}}_q(x_i)/2$.

Secondly, we add a term similar to eq.~(\ref{eq:rho_aa'}) where we
replace one of the non-perturbative 3-quark state of the proton by the
${\cal O}(g^2)$ virtual correction due to the exchange of a gluon by
two quarks, eq.~(\ref{eq:<x_i,r_i|R=0>_g2}). We also trace over the
quark colors. In all,
\bea
\rho^{qqq}_{\alpha\alpha'} &=& \left[1 - \frac{1}{2}\left(C^{\mathrm{reg}}_q(x_1) +
  C^{\mathrm{reg}}_q(x_2) +C^{\mathrm{reg}}_q(x_3) +C^{\mathrm{reg}}_q(x_1')
  +C^{\mathrm{reg}}_q(x_2') +C^{\mathrm{reg}}_q(x_3')\right)\right]\,\,
\Psi_\mathrm{qqq}^*(x_i', \vec r_i\!')\,\, \Psi_\mathrm{qqq}(x_i, \vec r_i)\nn\\
& & +
2g^2 C_F\,\int[\dd^2 q_i]\, 
\int_{\Delta x} \frac{\dd x_g}{2x_g}\frac{\dd^2 k_g}{(2\pi)^3}\,
\frac{k_g^2}{(k_g^2+\Delta^2)^2}\nn\\
& & \left\{
e^{i\sum\vec q_i\cdot\vec r_i}\,
\Psi_\mathrm{qqq}^*(x_i', \vec r_i\!')\,
\left[
\Psi_\mathrm{qqq}\left(x_1,\vec q_1+\vec k_g; x_2,\vec q_2-\vec k_g;
x_3, \vec q_3\right)
+
\Psi_\mathrm{qqq}\left(x_1,\vec q_1+\vec k_g; x_2,\vec q_2;
x_3, \vec q_3-\vec k_g\right) \right. \right.\nn\\
& & \hspace{4cm}\left.
+
\Psi_\mathrm{qqq}\left(x_1,\vec q_1; x_2,\vec q_2+\vec k_g;
x_3, \vec q_3-\vec k_g\right)\right] \nn\\
& & +
e^{-i\sum\vec q_i\cdot\vec r_i\!'}\,
\Psi_\mathrm{qqq}(x_i, \vec r_i)\,
\left[
\Psi^*_\mathrm{qqq}\left(x_1',\vec q_1+\vec k_g; x_2',\vec q_2-\vec k_g;
x_3', \vec q_3\right)
+
\Psi^*_\mathrm{qqq}\left(x_1',\vec q_1+\vec k_g; x_2',\vec q_2;
x_3', \vec q_3-\vec k_g\right) \right. \nn\\
& & \hspace{4cm}\left.\left.
+
\Psi^*_\mathrm{qqq}\left(x_1',\vec q_1; x_2',\vec q_2+\vec k_g;
x_3', \vec q_3-\vec k_g\right)\right]\right\}~.
\label{eq:rho_qqq_virt} \\
&=& \left[1 - \frac{1}{2}\left(C^{\mathrm{reg}}_q(x_1) +
  C^{\mathrm{reg}}_q(x_2) +C^{\mathrm{reg}}_q(x_3) +C^{\mathrm{reg}}_q(x_1')
  +C^{\mathrm{reg}}_q(x_2') +C^{\mathrm{reg}}_q(x_3')\right)\right]\,\,
\Psi_\mathrm{qqq}^*(x_i', \vec r_i\!')\,\, \Psi_\mathrm{qqq}(x_i, \vec r_i)\nn\\
& & +
2g^2 C_F\,
\Psi_\mathrm{qqq}^*(x_i', \vec r_i\!')\,\Psi_\mathrm{qqq}(x_i, \vec r_i)\,
\int_{\Delta x} \frac{\dd x_g}{2x_g}\frac{\dd^2 k_g}{(2\pi)^3}\,
\frac{k_g^2}{(k_g^2+\Delta^2)^2} \nn\\
& &~~~~
\left[
e^{-i\vec k_g\cdot(\vec r_1-\vec r_2)}\,
+
e^{-i\vec k_g\cdot(\vec r_1-\vec r_3)}
+
e^{-i\vec k_g\cdot(\vec r_2-\vec r_3)}
+
e^{i\vec k_g\cdot(\vec r_1\!'-\vec r_2\!')}\,
+
e^{i\vec k_g\cdot(\vec r_1\!'-\vec r_3\!')}\,
+
e^{i\vec k_g\cdot(\vec r_2\!'-\vec r_3\!')}\,
\right]~.
\label{eq:rho_qqq_virt_b}
\eea
Here, as in eq.~(\ref{eq:rho_r_r'_coord}), $\alpha=\{x_i, \vec r_i|\,
\sum x_i=1, \sum x_i\vec r_i=0\}$ and $\alpha'=\{x_i', \vec r_i\!'|\,
\sum x_i'=1, \sum x_i'\vec r_i\!'=0\}$ denote two sets of quark LC momentum
fractions and transverse positions.

Now we proceed  to $\rho^{qqqg}$.  We trace it over quark and gluon colors, and, in
addition, for simplicity over the gluon polarizations. Using
eq.~(\ref{eq:<alpha|R=0>_g}) in the definition~(\ref{eq:rho_aa'}) we
obtain
\bea
\rho^{qqqg}_{\alpha\alpha'} &=& 2g^2C_F \,
\int[\dd^2 k_i]\, [\dd^2 k_i']\,\,
e^{i\sum\vec k_i\cdot\vec r_i-i\sum\vec k_i'\cdot\vec r_i\!'}\,\,
\frac{\vec k_g\cdot\vec k_g'}{(k_g^2+\Delta^2)\,(k_g^{\prime 2}+\Delta^2)}
\nn\\
& &
\left\{2 \left(
\Psi^*_\mathrm{qqq}(k_1'+k_g'; k_2'; k_3')\,\Psi_\mathrm{qqq}(k_1+k_g; k_2; k_3)
+
\Psi^*_\mathrm{qqq}(k_1';k_2'+k_g'; k_3')\,\Psi_\mathrm{qqq}(k_1;k_2+k_g; k_3)
\right.\right. \nn\\
& &~~~~~~~~~~~~~~~~~\left.\left.
+
\Psi^*_\mathrm{qqq}(k_1'; k_2'; k_3'+k_g')\, \Psi_\mathrm{qqq}(k_1; k_2; k_3+k_g)
\right) \right. \label{eq:rho_qqqg_aa'-line1}
\nonumber\\
& &
- \Psi_\mathrm{qqq}(k_1+k_g; k_2; k_3)\,\Psi^*_\mathrm{qqq}(k_1';k_2'+k_g'; k_3')
- \Psi_\mathrm{qqq}(k_1+k_g; k_2; k_3)\,\Psi^*_\mathrm{qqq}(k_1';k_2'; k_3'+k_g')
\nn\\
& &
- \Psi_\mathrm{qqq}(k_1; k_2+k_g; k_3)\,\Psi^*_\mathrm{qqq}(k_1'+k_g';k_2'; k_3')
- \Psi_\mathrm{qqq}(k_1; k_2+k_g; k_3)\,\Psi^*_\mathrm{qqq}(k_1';k_2'; k_3'+k_g')
\nn\\
& & \left.
- \Psi_\mathrm{qqq}(k_1; k_2; k_3+k_g)\,\Psi^*_\mathrm{qqq}(k_1'+k_g';k_2'; k_3')
- \Psi_\mathrm{qqq}(k_1; k_2; k_3+k_g)\,\Psi^*_\mathrm{qqq}(k_1';k_2'+k_g'; k_3')
\right\}~.    \label{eq:rho_qqqg_aa'-line2}
\eea
Here, $\alpha=\{x_i, \vec r_i|\, \sum x_i=1, \sum x_i\vec r_i=0\}$ and
$\alpha'=\{x_i', \vec r_i\!'|\, \sum x_i'=1, \sum x_i'\vec r_i\!'=0\}$
denote two sets of quark and gluon momentum fractions and
transverse positions.

In Appendix~\ref{sec:app-traces} we show that the density matrix is
indeed properly normalized.

\section{Entanglement entropy of the perturbative density matrix}
\label{sec:EEpert-rho}

In this section we calculate the entanglement entropy of the density
matrix which includes one perturbatively emitted gluon.  We will
change our strategy somewhat to simplify the calculation. Integrating all
degrees of freedom in $A$ and calculating the entanglement entropy
turns out to be rather awkward as there are many degrees of freedom in
$\bar A$. Instead we choose to reduce the density matrix to a partial
set of degrees of freedom in the whole proton wave function, and only
then we integrate over $A$. We will follow two different routes.

In subsection~\ref{sec:S-O(g2)-q} we reduce the density matrix
calculated above by tracing over the gluon degrees of freedom in the
whole space. The resulting quark density matrix is then traced over
$A$ and the associated entanglement entropy is calculated. Note that
already after integrating over the gluon degrees of freedom the quark
density matrix does not describe a pure state and therefore in all
probability carries a nonvanishing entropy (which we do not calculate
here). Thus the entropy we calculate is not exactly the entanglement
entropy between the two spatial regions $A$ and $\overline A$, but
instead measures entanglement of quarks in $\overline A$ with the rest
of the proton wave function\footnote{Quantum correlations of regions
  $A$ and $\overline A$ could be analyzed using entanglement measures
  other than the von~Neumann entropy, which apply also to mixed
  states.  One such example is entanglement
  negativity~\cite{Vidal:2002q,Plenio:2005qn} which has been used
  recently to study two-quark azimuthal correlations in the
  light-cone wave function of the proton~\cite{Dumitru:2023fih}.}
(quarks in $A$ and gluon anywhere).

In subsection~\ref{sec:S-O(g2)-g} we perform a complementary
procedure: we integrate over the quark degrees of freedom in the whole
space, and then reduce the resulting gluon density matrix over $A$ and
calculate the entanglement entropy. Again, this entropy measures
entanglement of gluons in $\overline A$ with the rest of the proton
wave function.

\subsection{Entanglement entropy of quarks} \label{sec:S-O(g2)-q}
Let us construct the three-quark density matrix by tracing out the
gluon degrees of freedom in the whole space. Integrating over the
gluon leads to the density matrix:
\be
\rho \,=\, 
\rho^{qqq} + \tr_g\, \rho^{qqqg}\,.
\label{eq:rhoNLO-traced_g}
\ee
The first term, $\rho^{qqq}$ is given in eq.~(\ref{eq:rho_qqq_virt_b}).
To trace $\rho^{qqqg}_{\alpha\alpha'}$ over the gluon we  set $\vec r_g\!' = \vec r_g$ in
eq.~(\ref{eq:rho_qqqg_aa'-line1}), and
integrate with the measure $\dd x_g/(2x_g)\, 2\pi\dd^2 r_g$.  In
principle, the upper limit of $x_g$ in each term
of~(\ref{eq:rho_qqqg_aa'-line1}) is
different. However, in the small-$x_g$ approximation which 
we are employing here, only the leading $\log 1/x$ contribution
is important and we may replace the upper limits by a typical quark momentum
fraction $\langle x_q\rangle$. We then obtain
\bea
\tr_g\, \rho^{qqqg}_{\alpha\alpha'} &=& 2g^2C_F \,
\int_{\Delta x}\frac{\dd x_g}{2x_g}\int\frac{\dd^2 k_g}{(2\pi)^3}\,
\frac{1}{k_g^2+\Delta^2}
\int[\dd^2 k_i]\, [\dd^2 k_i']\,\,
e^{i\sum\vec k_i\cdot\vec r_i-i\sum\vec k_i'\cdot\vec r_i\!'}\,\,
\Psi^*_\mathrm{qqq}(k_i')\,\Psi_\mathrm{qqq}(k_i)
\nn\\
& &
\left\{2 \left(
e^{-i\vec k_g\cdot(\vec r_1-\vec r_1\!')}\,
+
e^{-i\vec k_g\cdot(\vec r_2-\vec r_2\!')}\,
+
e^{-i\vec k_g\cdot(\vec r_3-\vec r_3\!')}\,
\right) \right. \nn\\
& & \left.
- e^{-i\vec k_g\cdot(\vec r_1-\vec r_2\!')}\,
- e^{-i\vec k_g\cdot(\vec r_1-\vec r_3\!')}\,
- e^{-i\vec k_g\cdot(\vec r_2-\vec r_1\!')}\,
- e^{-i\vec k_g\cdot(\vec r_2-\vec r_3\!')}\,
- e^{-i\vec k_g\cdot(\vec r_3-\vec r_1\!')}\,
- e^{-i\vec k_g\cdot(\vec r_3-\vec r_2\!')}\,
\right\}~.
\eea
This can be written in terms of the position space wave functions
(\ref{eq:Psi_position_space}),
\bea
\tr_g\, \rho^{qqqg}_{\alpha\alpha'} &=& 2g^2C_F \,
\Psi^*_\mathrm{qqq}(x_i',\vec r_i\!')\,\Psi_\mathrm{qqq}(x_i,\vec r_i)
\int_{\Delta x}\frac{\dd x_g}{2x_g}\int\frac{\dd^2 k_g}{(2\pi)^3}\,\left\{
\left(\frac{1}{k_g^2+\Delta^2}-\frac{1}{k_g^2+\Lambda^2}\right)
\right. \\
& & \hspace{4cm}
2 \left(
e^{-i\vec k_g\cdot(\vec r_1-\vec r_1\!')}\,
+
e^{-i\vec k_g\cdot(\vec r_2-\vec r_2\!')}\,
+
e^{-i\vec k_g\cdot(\vec r_3-\vec r_3\!')}\,
\right) \nn\\
& & \left. -
\frac{1}{k_g^2+\Delta^2}\left[
e^{-i\vec k_g\cdot(\vec r_1-\vec r_2\!')}\,
+e^{-i\vec k_g\cdot(\vec r_1-\vec r_3\!')}\,
+e^{-i\vec k_g\cdot(\vec r_2-\vec r_1\!')}\,
+e^{-i\vec k_g\cdot(\vec r_2-\vec r_3\!')}\,
+e^{-i\vec k_g\cdot(\vec r_3-\vec r_1\!')}\,
+e^{-i\vec k_g\cdot(\vec r_3-\vec r_2\!')}
\right]
\right\}~.\nonumber
\eea
where we have reinstated the UV regulator $\Lambda$~\footnote{The
  dependence on the IR cutoffs $\Delta x$ and $\Delta^2$, and on the UV
  regulator $\Lambda^2$ cancels when eq.~(\ref{eq:rhoNLO-traced_g}) is
  traced over the quark degrees of freedom, as shown in
  Appendix~\ref{sec:app-traces}\,.}.\\

Let us now discuss the entropy of the density
matrix~(\ref{eq:rhoNLO-traced_g}).  Both
terms in~(\ref{eq:rhoNLO-traced_g}) are proportional to the
LO density matrix $\rho^\mathrm{LO}_{\alpha\alpha'} =
\Psi^*_\mathrm{qqq}(x_i', \vec r_i\!')\,\, \Psi_\mathrm{qqq}(x_i, \vec
r_i)$ discussed in sec.~\ref{sec:qqqDM-LO}:
\be
\rho^\mathrm{qqq}_{\alpha\alpha'} = F(\alpha,\alpha')\,
\rho^\mathrm{LO}_{\alpha\alpha'}~~~~~,~~~~~
\tr_g\, \rho^\mathrm{qqqg}_{\alpha\alpha'} = G(\alpha,\alpha')\,
\rho^\mathrm{LO}_{\alpha\alpha'}
\ee
with
\bea
F(\alpha,\alpha') &=&
1 - 3C^{\mathrm{reg}}_q(\langle x_q\rangle)
  + 2g^2 C_F\,
\int_{\Delta x} \frac{\dd x_g}{2x_g}\int\frac{\dd^2 k_g}{(2\pi)^3}\,
\frac{1}{k_g^2+\Delta^2} \nn\\
& & \hspace{1.5cm} \left\{
e^{-i\vec k_g\cdot(\vec r_1-\vec r_2)}\,
+
e^{-i\vec k_g\cdot(\vec r_1-\vec r_3)}
+
e^{-i\vec k_g\cdot(\vec r_2-\vec r_3)}
+
e^{i\vec k_g\cdot(\vec r_1\!'-\vec r_2\!')}\,
+
e^{i\vec k_g\cdot(\vec r_1\!'-\vec r_3\!')}\,
+
e^{i\vec k_g\cdot(\vec r_2\!'-\vec r_3\!')}\,
\right\}
\nn\\
&=& 1 - 3C^{\mathrm{reg}}_q(\langle x_q\rangle)
+ \frac{2g^2 C_F}{4\pi^2}\,\int_{\Delta x} \frac{\dd x_g}{2x_g}
\left\{
K_0(|\vec r_1-\vec r_2|\, \Delta)
+ K_0(|\vec r_1-\vec r_3|\, \Delta)
+ K_0(|\vec r_2-\vec r_3|\, \Delta)  \right. \nn\\
& & \left.~~~~~~~~~~~~~~~~
+ K_0(|\vec r_1\!'-\vec r_2\!'|\, \Delta)
+ K_0(|\vec r_1\!'-\vec r_3\!'|\, \Delta)
+ K_0(|\vec r_2\!'-\vec r_3\!'|\, \Delta)
\right\} \nn\\
G(\alpha,\alpha') &=&
2g^2C_F \,
\int_{\Delta x}\frac{\dd x_g}{2x_g}\int\frac{\dd^2 k_g}{(2\pi)^3}\,
\left\{
\left(\frac{1}{k_g^2+\Delta^2}-\frac{1}{k_g^2+\Lambda^2}\right)
\,\, 2 \left(
e^{-i\vec k_g\cdot(\vec r_1-\vec r_1\!')}\,
+
e^{-i\vec k_g\cdot(\vec r_2-\vec r_2\!')}\,
+
e^{-i\vec k_g\cdot(\vec r_3-\vec r_3\!')}\,
\right) \right. \nn\\
& & \left. - \frac{1}{k_g^2+\Delta^2}\left[
e^{-i\vec k_g\cdot(\vec r_1-\vec r_2\!')}\,
+e^{-i\vec k_g\cdot(\vec r_1-\vec r_3\!')}\,
+e^{-i\vec k_g\cdot(\vec r_2-\vec r_1\!')}\,
+e^{-i\vec k_g\cdot(\vec r_2-\vec r_3\!')}\,
+e^{-i\vec k_g\cdot(\vec r_3-\vec r_1\!')}\,
+e^{-i\vec k_g\cdot(\vec r_3-\vec r_2\!')}\right]
\right\} \nn\\
&=&
\frac{2g^2C_F}{4\pi^2} \int_{\Delta x}\frac{\dd x_g}{2x_g}
\, \left\{
  2[K_0(|\vec r_1-\vec r_1\!'|\, \Delta)-  K_0(|\vec r_1-\vec r_1\!'|\, \Lambda)]
+  2[K_0(|\vec r_2-\vec r_2\!'|\, \Delta)-  K_0(|\vec r_2-\vec r_2\!'|\, \Lambda)]
\right.\nn\\
& & ~~~~~~~~~~~~~~~
+  2[K_0(|\vec r_3-\vec r_3\!'|\, \Delta)-  K_0(|\vec r_3-\vec r_3\!'|\, \Lambda)]
  \nn\\
  & & ~~~~~~\left.
-K_0(|\vec r_1-\vec r_2\!'|\, \Delta)
-K_0(|\vec r_1-\vec r_3\!'|\, \Delta)
-K_0(|\vec r_2-\vec r_1\!'|\, \Delta) \right.\nn\\
& & ~~~~~~~~~~~~~~~~~~~~\left.
-K_0(|\vec r_2-\vec r_3\!'|\, \Delta)
-K_0(|\vec r_3-\vec r_1\!'|\, \Delta)
-K_0(|\vec r_3-\vec r_2\!'|\, \Delta)
\right\}
~.   \label{eq:rho_qqq-O_g2-G}
\eea
Interestingly, the diagonal matrix elements are unaffected by the presence of the gluon in the wave function, since $F(\alpha,\alpha)+G(\alpha,\alpha)=1$ due to real-virtual cancellations. Also note that integration over the gluon reinstates  the center of mass constraint for the coordinates of the three quarks. 
\\

After tracing over region $A$ both terms become block-diagonal in the
quark number basis, since the integration over the gluon results in a
reduced density matrix with fixed number of particles. The sub-blocks
correspond to 0, 1, 2, 3 quarks in $\overline A$, like in
sec.~\ref{sec:rho_Abar-qqq}:
\be
\tr_{A} \, \rho^{qqq} \,=\, 
\begin{pmatrix}
\rho_0^{(F)} & 0 & 0 & 0\\
0 & \rho_1^{(F)} & 0 & 0 \\
0 & 0 & \rho_2^{(F)} & 0 \\
0 & 0 & 0 & \rho_3^{(F)} \\
\end{pmatrix}
~~~~,~~~~
\tr_{A} \, \tr_g\, \rho^{qqqg} \,=\, 
\begin{pmatrix}
\rho_0^{(G)} & 0 & 0 & 0\\
0 & \rho_1^{(G)} & 0 & 0 \\
0 & 0 & \rho_2^{(G)} & 0 \\
0 & 0 & 0 & \rho_3^{(G)} \\
\end{pmatrix}\,.
\ee
Recall from sec.~\ref{sec:rho_Abar-qqq} that the $\rho_2$ and $\rho_3$
matrices are not diagonal in coordinate space (and their off-diagonal
elements do get modified at ${\cal O}(g^2)$) but that $\rho_1$ is, due to
the COM constraint.  In the limit of small $L$, $\rho_0$ and $\rho_1$
give the leading contribution $\sim L^2$ to the entropy.

For $\rho_0$ all quarks are in the region $A$ that we trace over, so
$\vec r_i = \vec r_i\!'$ and only the diagonal matrix elements of the
density matrix \eqref{eq:rhoNLO-traced_g} contribute. Since
$F(\alpha,\alpha)+G(\alpha,\alpha)=1$, the constant $\rho_0$ remains
equal to its value at LO, for any $L$. Hence, the derivative of
$S^{(0)}$ for $L\to0$ remains $\partial S^{(0)}/\partial L \sim L$
with the same numerical coefficient as in
eq.~(\ref{eq:dS0/dL_coefficient}).  \\

Now we consider $\rho_1$. Since this block is diagonal in the quark indices, we only need to consider
\be
(\rho_1)_{\alpha\alpha} =
3\Delta x\,a^2 \int \frac{\dd x_1\dd x_2}{8 x_1 x_2 x_3}\,
\delta(1-\sum x_i)
\int \dd^2r_1\, \dd^2r_2\, \delta(\sum x_i\vec r_i)\,
\Theta_A(\vec r_1)\,\Theta_A(\vec r_2)\,\,
[F(\vec r_i)+G(\vec r_i)]\,\,|\Psi(x_i,\vec r_i)|^2
~.
\label{eq:rho1_F}
\ee
Here $\alpha=\{x_3, \vec r_3\in \overline A\}$.
The perturbative correction again cancels as the sum
$F(\vec r_i)+G(\vec r_i)=1$, and we return to the expression from
sec.~\ref{sec:rho_Abar-qqq}.  The trace [$\int(\dd x_3/\Delta x)\,
(\dd^2r_3/a^2)\,
\Theta_{\overline A}(\vec r_3)$] vanishes at $L=0$ so there is no
contribution to $S(L=0)$. For $L>0$,
\bea
S^{(1)} &=& -
3\int[\dd x_i]\,[\dd^2 r_i]\, \Theta_{\overline A}(\vec r_3)\,
\Theta_{A}(\vec r_1)\,\Theta_{A}(\vec r_2)\,
\left[F(\vec r_i)+G(\vec r_i)\right]\, |\Psi(x_i,\vec r_i)|^2 \nn\\
& & \hspace{2cm}\log\left( 3\Delta x\, a^2
\int[\dd y_i]\,[\dd^2 s_i]\, \delta(\vec s_3-\vec r_3)\,
\delta(x_3-y_3)\,
\Theta_{A}(\vec s_1)\,\Theta_{A}(\vec s_2)\,
|\Psi(y_i,\vec s_i)|^2\right)~.
\eea
The derivative of $S^{(1)}$ w.r.t.\ $L$ for $L\to0$ is
proportional to $L$ with the coefficient given in
eq.~(\ref{eq:dS1/dL_coefficient}).
\\

To summarize, we find that due to real-virtual cancellations in gluon
emission, the leading (at small $L$) term in the entanglement entropy of quarks
is identical to that for the initial non-perturbative three-quark wave
function.\\

Let us now take a look at $\rho_2$ (two quarks inside the circle
separated by a typical distance of order $L$).  It is given by the LO
expression, eq.~(\ref{eq:rho_2}), times $F(\vec r_1,\vec r_2,\vec r_3;
\vec r_1\!',\vec r_2\!',\vec r_3) + G(\vec r_1,\vec r_2,\vec r_3; \vec
r_1',\vec r_2\!',\vec r_3)$.  Due to the COM constraint $x_1\vec r_1 +
x_2\vec r_2 = x_1'\vec r_1\!' + x_2'\vec r_2\!' = - x_3\vec r_3$, and
$x_1+x_2= x_1'+x_2'=1-x_3$, so that there is in fact no integral over
the coordinates of the quark in $A$ ($\vec r_3$ or $x_3$) as those are
completely determined by the coordinates and momentum fractions of the
two quarks in $\overline A$.

We need $F+G$ for $\vec r_3 = \vec r_3\!'$, and we shall use the
position space (Bessel function) form of these functions from
eq.~(\ref{eq:rho_qqq-O_g2-G}).
In the limit $\vec r_3\!' \to \vec r_3$ we have that
$2K_0(|\vec r_3-\vec r_3\!'|\, \Delta)-  2K_0(|\vec r_3-\vec r_3\!'|\, \Lambda)
\to \log \frac{\Lambda^2}{\Delta^2}$.
Furthermore, we consider $K_0(|\vec r_1-\vec r_1\!'|\, \Lambda)$ and
$K_0(|\vec r_2-\vec r_2\!'|\, \Lambda)$  to be exponentially
small since generically $|\vec r_1-\vec r_1\!'|, |\vec r_2-\vec r_2\!'|
\sim L$ and $L\Lambda\gg1$. With that, and noting that $L\Delta\ll1$,
we can write
\bea
F+G &\simeq &
1 - \frac{2g^2C_F}{4\pi^2} \, \log\frac{x_q}{\Delta x}\,
\log\frac{\Lambda^2}{\Delta^2}
+ \frac{g^2 C_F}{8\pi^2}\,\int_{\Delta x} \frac{\dd x_g}{x_g}
\left[
  \log\frac{1}{(\vec r_1-\vec r_2)^2\, \Delta^2}
 +\log\frac{1}{(\vec r_1\!'-\vec r_2\!')^2\, \Delta^2}
 +2\log\frac{1}{(\vec r_1-\vec r_1\!')^2\, \Delta^2} \right.\nn\\
 & & ~~~~~~~~~~~~~~~~~~~~~~~~~~~~~\left.
 +2\log\frac{1}{(\vec r_2-\vec r_2\!')^2\, \Delta^2}
 -\log\frac{1}{(\vec r_1-\vec r_2\!')^2\, \Delta^2}
 -\log\frac{1}{(\vec r_2-\vec r_1\!')^2\, \Delta^2}
 \right] \\
& \simeq &
1 - \frac{2g^2C_F}{4\pi^2} \, \log\frac{x_q}{\Delta x}\, \log L^2\Lambda^2 +
\frac{g^2 C_F}{8\pi^2}\,\log\frac{x_q}{\Delta x}\,
  \log\frac{(\vec r_1-\vec r_2\!')^2\,(\vec r_2-\vec r_1\!')^2}
           {(\vec r_1-\vec r_2)^2\,(\vec r_1\!'-\vec r_2\!')^2} \\
& \simeq &
1 - \frac{2g^2C_F}{4\pi^2} \, \log\frac{x_q}{\Delta x}\, \log L^2\Lambda^2
~.
\eea
The equality here is valid with leading logarithmic accuracy, since in
the second step, $\log\frac{1}{(\vec r_1-\vec r_1\!')^2\, \Delta^2}$ and
$\log\frac{1}{(\vec r_2-\vec r_2\!')^2\, \Delta^2}$ were replaced by
$\log\frac{1}{L^2\Delta^2}$; and in the last step an $O(1)$
(non-logarithmic) term was dropped.
\begin{figure}[htb]
  \includegraphics[width=0.38\textwidth]{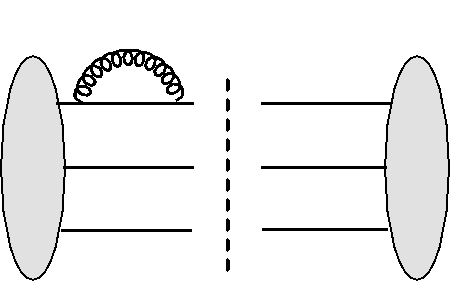} \hspace*{1.5cm}
  \includegraphics[width=0.38\textwidth]{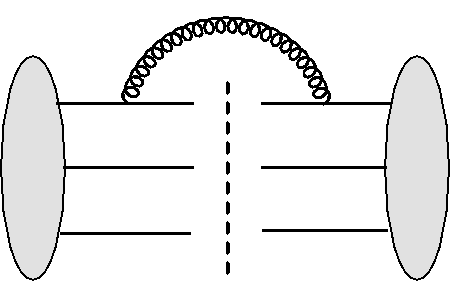}
  \caption{\label{fig:qqDM-diagram} Left (virtual correction):
    the transverse position of the gluon
    emitting and gluon absorbing quark is the same; hence
    here the gluon transverse momentum is integrated up to
    $\Lambda$, and we obtain a contribution $\sim \log
    \frac{\Lambda^2}{\Delta^2}$.\\
    Right (real emission): here there is a mismatch of order $L$ in
    the transverse positions of quarks 1, 1' across the cut, and the
    diagram is only $\sim \log\frac{1}{L^2\Delta^2}$.
  }
\end{figure}

With these simplifications $F+G$ is just a number, i.e.\ it does not
modify the matrix structure of $\rho_2$ relative to the leading order,
but only multiplies the entire matrix by a numerical prefactor.  Its
contribution to the entropy $S^{(2)}$ is given by the last term in
eq.~(\ref{eq:SvN}) with the substitution $|\Psi(x_i,\vec r_i)|^2 \to
(F+G)\, |\Psi(x_i,\vec r_i)|^2$:
\bea\label{s2}
S^{(2)} &=& -
3 \left(1 - \frac{2 g^2 C_F}{4\pi^2} \log\frac{x_q}{\Delta x}\,
\log L^2\Lambda^2\right)
\int[\dd x_i]\,[\dd^2 r_i]\, \Theta_{A}(\vec r_3)\,
\Theta_{\overline A}(\vec r_1)\,\Theta_{\overline A}(\vec r_2)\,
|\Psi(x_i,\vec r_i)|^2\nn\\
& & \hspace{2cm} \log\left( 3\Delta x\, a^2
\int[\dd y_i]\,[\dd^2 s_i]\, \delta(\vec s_3-\vec r_3)\,
\delta(x_3-y_3)\,
\Theta_{\overline A}(\vec s_1)\,\Theta_{\overline A}(\vec s_2)\,
|\Psi(y_i,\vec s_i)|^2\right)
 ~.
\eea
Here, with leading logarithmic accuracy we have omitted the factor
$F+G$ under the logarithm.

Note that, as opposed to the entropy associated with $\rho_1$, this
contribution to the entropy does receive corrections from gluon
emission. This arises because the two quarks may be at different
positions in the amplitude (1, 2) and the conjugate amplitude (1',
2'). The mismatch between these positions is generically of order $L$,
c.f.\ Fig.~\ref{fig:qqDM-diagram}. For such configurations the
contribution of the ``real" diagrams, i.e.\ the diagrams where a quark
exchanges a gluon with itself across the cut is proportional to
$\log\frac{1}{L^2\Delta^2}$ rather than
$\log\frac{\Lambda^2}{\Delta^2}$ as is the case for the ``virtual"
diagrams (where the gluon is exchanged in the amplitude or the complex
conjugate amplitude).  Thus the real-virtual cancellation is
incomplete and leads to the logarithmic correction in \eqref{s2}. The
real-virtual cancellation essentially only occurs when the gluon is
emitted outside $\bar A$ but not inside $\bar A$.  Also note that the
perturbative correction to $S^{(2)}$ is negative, suggesting stronger
correlations between quarks when perturbative gluon emission is
included.

We note again, that $S^{(2)}$ is subleading for small $L^2$, and thus
only provides a small correction to the quark entanglement
entropy.

\subsection{Entanglement entropy of the gluon}
\label{sec:S-O(g2)-g}

We now integrate over the quark degrees of freedom in the whole
space. The resulting density matrix for the gluon has the general
structure
\be
\rho \,=\, 
\begin{pmatrix}
\tr_\mathrm{qqq}\, \rho^{qqq} & 0 \\
0 & \rho^{g} \\
\end{pmatrix}\,.
\label{eq:rhoNLO-traced_qqq}
\ee
The first block is just a number which is equal to the probability
that no gluons are present in the wave function. It is given by
the integral of the diagonal of eq.~(\ref{eq:rho_qqq_virt_b}) over $[\dd
  x_i]$ and $[\dd^2 r_i]$:
\bea
\tr_\mathrm{qqq}\, \rho^{qqq} &=& 1 -3C^{\mathrm{reg}}_q(\langle x_q\rangle)
+
4g^2 C_F\,\int[\dd x_i]\int[\dd^2 r_i]\,\,\left|
\Psi_\mathrm{qqq}(x_i, \vec r_i)\right|^2
\int_{\Delta x} \frac{\dd x_g}{2x_g}\int\frac{\dd^2 k_g}{(2\pi)^3}\,
\frac{1}{k_g^2+\Delta^2}\nn\\
& & ~~~~~~~~~~~~~~~~~~~~
\left[\cos \vec k_g\cdot(\vec r_1-\vec r_2) +
  \cos \vec k_g\cdot(\vec r_1-\vec r_3) + \cos
  \vec k_g\cdot(\vec r_2-\vec r_3)
  \right]~.   \label{eq:tr_qqq_O(g2)}
\eea

For the second block we return to eq.~(\ref{eq:rho_qqqg_aa'-line2})
and integrate the diagonal in the three-quark space ($x_i'=x_i$ and
$\vec r_i\!' = \vec r_i$ for the three quarks) over $[\dd x_i]$ and
$[\dd^2 r_i]$\footnote{ In principle these are integrations over the
  LC momentum fractions and transverse coordinates of the three quarks
  with COM constraints which include the gluon, since the density
  matrix in eqs.~(\ref{eq:rho_qqqg_aa'-line2}) was defined over the domain
  $\sum_{i=1}^4x_i=1$ , $\sum_{i=1}^4x_i\vec r_i=0$ .  However, when
  $x_g$ is very small, the presence of the gluon will not
  significantly restrict the integrations over quark $x_i, \vec r_i$,
  and we can approximate $x_g\approx 0$ in the $\delta$-functions for
  the COM constraints.}. Note that since we have traced out the
quarks in the whole space, the COM constraint forces $\vec r_g = \vec
r_g'$, and the gluon density matrix is diagonal:
\bea
\rho^{g}_{\alpha\alpha} &=& 2g^2C_F \,
\int\frac{\dd^2 k_g}{(2\pi)^3}
\int\frac{\dd^2 k_g'}{(2\pi)^3}
\frac{\vec k_g\cdot\vec k_g'}{(k_g^2+\Delta^2)\,(k_g^{\prime 2}+\Delta^2)}
\, e^{i\vec k_g\cdot\vec r_g-i\vec k_g'\cdot\vec r_g}\,\,
\int[\dd x_i]\, [\dd^2 r_i]\,\,
|\Psi_\mathrm{qqq}(x_i,\vec r_i)|^2
\nn\\
& &
\left\{2 \left(
e^{i(\vec k_g'-\vec k_g)\cdot\vec r_1}
+
e^{i(\vec k_g'-\vec k_g)\cdot\vec r_2}
+
e^{i(\vec k_g'-\vec k_g)\cdot\vec r_3}
\right) \right.\nn\\
& & \left.
- e^{i\vec k_g'\cdot\vec r_2-i\vec k_g\cdot\vec r_1}
- e^{i\vec k_g'\cdot\vec r_3-i\vec k_g\cdot\vec r_1}
- e^{i\vec k_g'\cdot\vec r_1-i\vec k_g\cdot\vec r_2}
- e^{i\vec k_g'\cdot\vec r_3-i\vec k_g\cdot\vec r_2}
- e^{i\vec k_g'\cdot\vec r_1-i\vec k_g\cdot\vec r_3}
- e^{i\vec k_g'\cdot\vec r_2-i\vec k_g\cdot\vec r_3}
\right\}~.  \label{eq:rho_g-alpha/=alpha'}
\eea
Here, $\alpha=\{x_g,\vec r_g\}$ and $\alpha'=\{x_g',\vec r_g\!'\}$.  As
before, the gluon ``propagators" have to be regularized in the UV by
the Pauli-Villars regulator. This entails substituting
$\frac{1}{k_g^2+\Delta^2}\rightarrow
\frac{1}{k_g^2+\Delta^2}-\frac{1}{k_g^2+\Lambda^2}$, and the same for
$k_g'$. We can simplify the resulting expression somewhat, noting that
the UV divergence only resides in the first term in the curly brackets
in \eqref{eq:rho_g-alpha/=alpha'}, since in the second term both
integrations, over $\vec k_g$ and $\vec k_g'$, are already regulated by
the phase factors, while in the first term the phase factors only
regulate the integration over $k_g-k_g'$. Also, it is sufficient to
regulate only one of the propagators in the product to eliminate the
UV divergence, but this regularization has to be done symmetrically
between $\vec k_g$ and $\vec k_g'$. Thus, we substitute
\begin{equation}
\frac{1}{(k_g^2+\Delta^2)(k_g^{\prime 2}+\Delta^2)}\rightarrow\frac{1}{(k_g^2+\Delta^2)(k_g^{\prime 2}+\Delta^2)}-\frac{1}{2}\left[
 \frac{1}{(k_g^2+\Delta^2)(k_g^{\prime 2}+\Lambda^2)}+\frac{1}{(k_g^2+\Lambda^2)(k_g^{\prime 2}+\Delta^2)}\right]
\end{equation}
{\em in the first term} in the curly brackets of
\eqref{eq:rho_g-alpha/=alpha'}. We also use the symmetry of the quark
wave function and the integration measure to rename coordinates in
some terms.  The resulting UV regular density matrix then is
\bea
\rho^{g}_{\alpha\alpha} &=& 12g^2C_F \, \int\frac{\dd^2 k_g}{(2\pi)^3}
\int\frac{\dd^2 k_g'}{(2\pi)^3} \frac{\vec k_g\cdot\vec
  k_g'}{(k_g^2+\Delta^2)\,(k_g^{\prime 2}+\Delta^2)} \int[\dd x_i]\,
         [\dd^2 r_i]\,\, |\Psi_\mathrm{qqq}(x_i,\vec r_i)|^2 \nn\\
& &~~~~~~~~
         \left\{ e^{i(\vec k_g'-\vec k_g)\cdot(\vec r_1-\vec r_g)} -
         e^{i(\vec k_g-\vec k_g')\cdot\vec r_g+i\vec k_g'\cdot\vec
           r_2-i\vec k_g\cdot\vec r_1} \right\}\nn\\
& & -6g^2C_F \,
         \int\frac{\dd^2 k_g}{(2\pi)^3} \int\frac{\dd^2
           k_g'}{(2\pi)^3} \left[ \frac{\vec k_g\cdot\vec
             k_g'}{(k_g^2+\Delta^2)(k_g^{\prime
               2}+\Lambda^2)}+\frac{\vec k_g\cdot\vec
             k_g'}{(k_g^2+\Lambda^2)(k_g^{\prime 2}+\Delta^2)}\right]\nn\\
& & ~~~~~~~~~~~~~~
         \int[\dd x_i]\, [\dd^2 r_i]\,\, |\Psi_\mathrm{qqq}(x_i,\vec
         r_i)|^2 e^{i(\vec k_g'-\vec k_g)\cdot(\vec r_1-\vec r_g)}
         ~.  \label{eq:rho_g-alpha/=alpha}
\eea
Note that when calculating trace of $\rho$, the regulator simply adds
the term
\be
- 12g^2C_F \int\frac{\dd^2 k_g}{(2\pi)^3}\,\frac{1}{\vec k_g^2 + \Lambda^2}~,
\ee
which (up to powers of $\Delta^2/\Lambda^2$ ) cancels the similar term
that arises from the regulator in $C_q^\mathrm{reg}$ in
\eqref{eq:tr_qqq_O(g2)}. Thus, our Pauli-Villars regularization
preserves the trace of the density matrix.

On the other hand,  $\tr\, \rho^{g}$ by itself has the meaning of the probability to find one gluon the proton wave function. 
The trace ($2\pi\int\dd^2 r_g\, \int \dd x_g/(2x_g)$) is given by
\bea\label{rhog2} \tr\, \rho^{g} &=& \frac{6 g^2 C_F}{8\pi^2}\,
\log\frac{\langle x_q\rangle}{\Delta x}\left[
  \log\frac{\Lambda^2}{\Delta^2} -2 \int[\dd x_i]\, [\dd^2 r_i]\,\,
  |\Psi_\mathrm{qqq}(x_i,\vec r_i)|^2\, K_0(|\vec r_2-\vec
  r_1|\,\Delta) \right]~
\eea
where $\Delta x$ as before is the IR cutoff on possible longitudinal
momenta and the integral over $x_g$ is cut off at $\langle x_q\rangle$
consistently with the soft gluon approximation.  In the second term
the integral is dominated by $|\vec r_2-\vec r_1|$ of order of the
collinear regulator $\Delta^{-1}$, so the second term is negligible.

Eq.~\eqref{rhog2} can be related to the gluon PDF of the proton.  To
leading order in perturbation theory (see for example eq.~(65a) in
Kovchegov and Mueller~\cite{Kovchegov:1998bi}):
\be
\frac{\alpha_s C_F}{\pi}\frac{1}{\ell^2} = \frac{\partial}
{\partial\ell^2}\, xG_q(x,\ell^2)~,
\ee
is the gluon PDF of a quark. Thus, for a proton consisting of three
quarks we identify the gluon PDF as
\be
\frac{3 g^2 C_F}{4\pi^2} \int \dd k^2\, \left[\frac{1}{k^2+\Delta^2}
  - \frac{1}{k^2+\Lambda^2}\right] =
\frac{3 g^2 C_F}{4\pi^2} \log\frac{\Lambda^2}{\Delta^2}
~ \to~  xG(x,\Lambda^2)~.
\ee
Hence, we have
\be \label{eq:tr-rho_g->PDF}
\tr\, \rho^{g} = \int\limits_{\Delta x}
\frac{\dd x_g}{x_g}\,\, x_g G(x_g,\Lambda^2)
= \int\limits_{\Delta x}\dd x_g\,\, G(x_g,\Lambda^2)~.
\ee
Indeed this is just the total number of gluons at the resolution scale
of the UV cutoff $\Lambda$. The fact that the UV cutoff appears in
this quantity is not surprising, since here we are dealing with the
density matrix of the entire proton wave function rather than the part
of it probed by a DIS probe. If we were to calculate the density
matrix of only those degrees of freedom that participate in a DIS process,
we expect that the UV cutoff would be substituted by the external
resolution scale $\Lambda^2\rightarrow Q^2$ provided by the virtual
photon.\\

Let us now construct the reduced density matrix after tracing over $A$. 
It is of the form
\be
\rho \,=\, 
\begin{pmatrix}
I + \rho_0^g & 0 \\
0 & \rho^g_1 \\
\end{pmatrix}~,
\label{eq:rhoNLO-traced_qqq-traced_A}
\ee
where $I\equiv \tr_\mathrm{qqq}\, \rho^{qqq}$ for brevity. The first entry is the probability that there are no gluons in $\bar A$, and is of course a pure number.
\be
\rho_0^g = \int_{\Delta x}\frac{\dd x_g}{2x_g}\,\, 2\pi\int\dd^2 r_g
\,\Theta_A(\vec r_g)\, \rho^g_{\alpha \alpha}~.
\ee

The lower block is a diagonal (in coordinate space) matrix 
\begin{equation}
  \rho_1^g(\vec r) =  \rho^g(\vec r)\Theta_{\bar A}(\vec r)~,
\end{equation}
with $\rho^g$ from eq.~\eqref{eq:rho_g-alpha/=alpha}.  Like for
quarks, we need to scale $\rho_1^g$ with the transverse-longitudinal
lattice spacing, and with the factor $2\pi/2x_g$ that accompanies the
integration measure $\dd x_g\, \dd^2 r_g$. We will not do it
explicitly here, but instead restore these factors directly in the
expression for the entropy.
For $L=0$, as already mentioned, $I + \rho_0^g=1$.  We are interested
in the non-trivial small-$L$ regime, $\Delta^{-1} \gg L \gg
\Lambda^{-1}$ or $L \Delta \ll 1 \ll \Lambda L$.  In this regime we
expect $I + \rho_0^g\sim 1- {\cal O}(\L^2)$. The contribution to the
entropy associated with this single eigenvalue of the density matrix
should be $S^{(0)}\sim {\cal O}(L^2)$. The matrix $\rho_1^g$ on the
other hand, has small eigenvalues, for the same reason discussed
previously. All the eigenvalues should be of order $(\Delta x)\,a^2
\,\Delta^2$, due to the dimensionality of $\rho_1^g$. Thus, we expect
the contribution from $\rho_1^g$ to the entropy to contain an additional
enhancement by a logarithm of $(\Delta x)\,a^2$:
\be  \label{eq:S^1}
S^{(1)} = - \int_{\Delta x}\frac{\dd x_g}{2x_g}\, 2\pi\int\dd^2 r_g\,\,
\Theta_{\overline A}(\vec r_g)\,\,
(\rho^g)_{\alpha \alpha}\,\log \left(2\pi a^2 \frac{\Delta x}{2x_g}\,
(\rho^g)_{\alpha \alpha}\right)~.
\ee
This therefore is the leading contribution to the entropy and we will calculate it first.

Let us examine 
 $(\rho^g)_{\alpha \alpha}$ for $|\vec r_g| <L
\ll \Delta^{-1}$. The first term in \eqref{eq:rho_g-alpha/=alpha}  is
\bea
& & 12 g^2 C_F\,
\int\frac{\dd^2 k_g}{(2\pi)^3}
\int\frac{\dd^2 q}{(2\pi)^3}
\left[
  \frac{1}{(\vec k_g+\vec q)^{2}+\Delta^2}
  +
  \frac{\vec k_g\cdot\vec q}{(\vec k_g^{2}+\Delta^2)\,
    ((\vec k_g+\vec q)^{2}+\Delta^2)}
  \right] \, e^{-i\vec q\cdot\vec r_g}\nn\\
& & ~~~~~\int[\dd x_i]\, [\dd^2 r_i]\,\,
|\Psi_\mathrm{qqq}(x_i,\vec r_i)|^2
\left\{
e^{i\vec q\cdot\vec r_1}
- e^{i\vec k_g\cdot(\vec r_2-\vec r_1)+i\vec q\cdot\vec r_2}
\right\} \,.
\eea
The integrals over the quark positions basically result in a ``smeared
$\delta$-function'' in $\vec q$ with width $\Delta$, so $|\vec
q|\sim\Delta$.  That means that the phase $e^{-i\vec q\cdot\vec
  r_g}\sim 1$ since $L\Delta\ll1$.  Furthermore, the denominator of
the second rational factor is essentially constant (independent of the
direction of $\vec q$) both for small and large $\vec k_g$; hence, it
gives zero after integration over the directions of $\vec q$. Lastly,
the second phase factor in the curly braces would restrict $|\vec
k_g|\sim\Delta$ which results in a subleading contribution.  In all,
we simplify the above to
\bea
& & 12 g^2 C_F\,
\int\frac{\dd^2 k_g}{(2\pi)^3}
  \frac{1}{\vec k_g^{2}+\Delta^2}
\int\frac{\dd^2 q}{(2\pi)^3}
\int[\dd x_i]\, [\dd^2 r_i]\,\,
|\Psi_\mathrm{qqq}(x_i,\vec r_i)|^2\, 
e^{i\vec q\cdot\vec r_1} \nn\\
&\simeq&
 12 g^2 C_F\,\frac{\Delta^2}{\pi}
\int\frac{\dd^2 k_g}{(2\pi)^4}
  \frac{1}{\vec k_g^{2}+\Delta^2}~.
\eea
We have assumed in these estimates that the nonperturbative scale
entering the gluon propagator ($\Delta^2$) and the nonperturbative
scale appearing in the quark wave function (the average quark
transverse momentum squared $\sim\beta^2$) are of the same order.

The second term in   \eqref{eq:rho_g-alpha/=alpha}  gives
a similar expression but with $\Delta$ replaced by $\Lambda$, and with the
opposite sign. In all, for $|\vec r_g| < L$,
\be\label{rhog}
(\rho^g)_{\alpha\alpha} \simeq
\frac{12 \, g^2 C_F}{(2\pi)^4}\, \Delta^2 \, \log\frac{\Lambda^2}{\Delta^2}~.
\ee
This has a simple interpretation. We are calculating the probability
density of a gluon to be emitted at point $\vec r$ inside $\bar
A$. Since the region inside $\bar A$ is small compared to the
proton ($L\ll 1/\Delta$) the
emission probability does not depend on $\vec r$. It is given (with
the appropriate prefactor) by the integral of the intensity of the
Weizs\"acker-Williams field of a quark, integrated over the coordinate of
the quark weighted with the square of the quark wave function. With
logarithmic accuracy this is simply $\int_{\Lambda^{-2}<r^2<\Delta^{-2}}
\dd^2r \frac{1}{r^2}$, which is precisely the
logarithm in \eqref{rhog}.

Using this in eq.~(\ref{eq:S^1}) we obtain
\bea
S^{(1)} &=& - L^2 \Delta^2 \,
\int_{\Delta x}\dd x_g\, G(x_g,\Lambda^2/\Delta^2)\,
\log\left( \frac{a^2\Delta^2}{\pi} ({\Delta x})\, G(x_g,\Lambda^2/\Delta^2)
\right)\nonumber\\
&=& - L^2 \Delta^2 \,
\int_{\Delta x}\dd x_g\, G(x_g,\Lambda^2/\Delta^2)\,
\log\left( \frac{\Delta^2}{\pi\Lambda^2} ({\Delta x})\, G(x_g,\Lambda^2/\Delta^2)
\right)~.\label{s1}
\eea
Once again, $(\Delta^2/\pi)\, G(x_g,\Lambda^2/\Delta^2)$ is the
density of gluons per unit transverse area.

Since the density matrix \eqref{eq:rhoNLO-traced_qqq-traced_A} is
normalized, we infer from \eqref{rhog}
\be
I+\rho_0^g = 1 - \frac{3g^2C_F}{4\pi^2}\,  L^2 \Delta^2 \,
\int_{\Delta x}\frac{\dd x_g}{x_g}\, \log \Lambda^2/\Delta^2~,
\ee
and the associated entropy is
\be
S^{(0)} = \frac{3g^2C_F}{4\pi^2}\,  L^2 \Delta^2 \,
\int_{\Delta x}\frac{\dd x_g}{x_g}\, \log \Lambda^2/\Delta^2
=L^2 \Delta^2 \,
\int_{\Delta x}\dd x_g\, G(x_g,\Lambda^2/\Delta^2)~.
\ee
This is the gluon density per unit transverse area multiplied by the
area of the cutout.  As expected, this is a subleading correction to
\eqref{s1} and can be neglected.

Thus our final result for the gluon entanglement entropy in the limit of small area of the cutout is given in eq.~\eqref{s1}.

\section{Discussion}  \label{sec:Discussion}

To summarize, we 
calculated the entanglement entropy of subsets (in several
variations) of partonic modes in the model proton wave function inside
a small disk of radius $L$ by integrating all the other modes in the
rest of the wave function. The area was taken small relative to the
total area of the proton (a soft, nonperturbative scale) $L^2\ll
\pi/\Delta^2$, but greater than the inverse UV cutoff
$L^2\gg1/\Lambda^2$.

We now want to comment on these results. Let us consider the two
expressions \eqref{sFqq} and \eqref{s1}. Eq.\ \eqref{sFqq} gives the
entanglement entropy of quarks at leading order in the model wave
function, while eq.~\eqref{s1} is the entanglement entropy of gluons
at NLO. They have almost identical structure and are reminiscent of
the form of Boltzmann entropy of a system of noninteracting
particles. The PDFs that enter \eqref{sFqq} and \eqref{s1} ($F$ in the
former and $G$ in the latter) are the total numbers of quarks and
gluons in the proton. Defining the number of partons (at a given $x$)
inside an area $S$, in the longitudinal momentum interval $\Delta x$,
as $N_S(x)=\frac{S}{A_p}(\Delta x)F(x)$ for quarks and
$N_S(x)=(S\Delta^2/\pi) \, (\Delta x)\, G(x)$ for gluons, both
equations can be written as 
\be\label{sb} S_E=-\int\frac{dx}{\Delta
  x}N_{L^2}(x)\log[N_{a^2}(x)]\,. \ee 
  This expression is quite
natural. For small $a^2$ and $\Delta x$, one can only have either one
or no partons inside the elementary cell $a^2\Delta x$. The average
number of partons $N_{a^2}(x)$ is then just the probability that the cell contains
 a single parton.  Eq.~\eqref{sb} then is just (the leading term of) the
Shannon entropy of this distribution multiplied by the total area (or
rather $L^2/a^2$ - the number of independent elementary cells in the
area of the cutout), and integrated over $x$ with the appropriate
measure.  The fact that the entropy is proportional to the area $L^2$
is a trademark property of an extensive quantity. Of course the
entanglement entropy is not strictly speaking extensive - the
proportionality to area only holds when the area of the cutout is
small. Were we to take the area of the cutout to be equal to the area
of the proton we would have to obtain vanishing entropy as we would
not be integrating out any degrees of freedom. So, the dependence of
entropy on area should follow some sort of a Page curve which could be
obtained numerically from eq.~\eqref{eq:SvN}, for example. 

 The one
significant difference between eqs.~\eqref{sFqq} and \eqref{s1} is
that in the latter the number of particles is defined with the
resolution scale $\Lambda^2$, as is appropriate in the QCD improved
parton model, while in the former there is no need to specify a
resolution scale.\\

Does the entropy calculated here have direct physical meaning? One
should remember, of course, that the calculation presented here does
not refer to any particular physical process, but rather to the
properties of the proton wave function {\it per se}. As such it is not
observable directly. We can try, however, to interpret this result
from the point of view of a DIS or jet production process. In this
type of process there is a physical resolution scale, the momentum
transfer $Q^2$ to the electron or the transverse momentum of a
produced jet. A naive physical picture is then that this scale should
determine the size of the area of the proton measured by the probe, as
well as the resolution with which one measures the parton
number. Taking $L^2 \sim a^2 \sim 1/Q^2$, $\Delta^2 = \pi
\Lambda_\mathrm{QCD}^2$, and fixing the value of $x$ as appropriate for
DIS, we then may hope to define a more physical quantity. For gluons
that would be
\begin{equation}\label{sbqx}
  S_E(Q^2,x) =
  - N_{Q^{-2}}(x)\log[N_{Q^{-2}}(x)] =
  - \frac{\Lambda^2_\mathrm{QCD}}{Q^2}\, (\Delta x)G(x,Q^2)
  \log \left(\frac{\Lambda^2_\mathrm{QCD}}{Q^2}(\Delta x)G(x,Q^2)\right)
\end{equation}
It is not entirely clear to us what should be taken as the
``longitudinal resolution scale" $\Delta x$. The inclusive DIS cross
section does not provide for a scale of this sort. However, if one
measures the spectrum of produced particles, perhaps $\Delta x$ should be
related to the width of the rapidity bin in which the particles are
measured.

Finally, it would be interesting to compare our results with those of
ref.~\cite{Kharzeev:2017qzs}. This may not be entirely straightforward
for the following reason. Our expressions apply to the ``dilute
regime" when the entropy is dominated by states with one parton within
the cutout area, $\frac{\Lambda^2_{\rm QCD}}{Q^2}(\Delta x)G(x,Q^2)\ll
1$. On the other hand ref.~\cite{Kharzeev:2017qzs} focused on the
saturation regime where the number of particles in the cutout is
assumed to be ${\cal O}(1/\alpha_s)$. Still, the actual derivation of
ref.~\cite{Kharzeev:2017qzs} only requires that the rapidity is large
enough so that the exponential growth of the gluon density in rapidity
has taken hold. This in itself does not imply saturation, but rather
the pre-saturation BFKL-like regime, so that the gluon density is
still small but low-$x$ evolution already has to be resummed.

At any rate, one expects the same elements to appear in the expression
for entropy both in ref.~\cite{Kharzeev:2017qzs} and in our
calculation. Indeed, the parton density is the basic physical quantity
that appears, and in this respect the two results are similar. However
there are some significant differences between the two. In particular,
according to ref.~\cite{Kharzeev:2017qzs} the entropy is given by the
logarithm of $xG(x)$. This is somewhat perplexing since $xG(x)$ has the
meaning of the longitudinal momentum carried by the partons, and not
the parton number. Eq.~\eqref{sbqx} on the other hand contains
$\frac{\Lambda^2_{\rm QCD}}{Q^2}(\Delta x)G(x,Q^2)$ which is precisely the
number of partons in the area of the cutout (and in the rapidity
interval $\Delta x$) , which appears to be the natural basic element
for quantifying the entropy. Whether the number of partons at high
energy is somehow supplanted by the longitudinal momentum fraction carried by the partons is an
interesting question which should be answered by an explicit
calculation.

\section*{Acknowledgements}

A.D.\ acknowledges support by the DOE Office of Nuclear Physics
through Grant DE-SC0002307. A.K. is supported by the NSF Nuclear
Theory grant 2208387.  V.S.\ is supported by the U.S. Department of
Energy, Office of Science, Office of Nuclear Physics through the
Contract No. DE-SC002008. We also acknowledge support by the Saturated
Glue (SURGE) Topical Collaboration.

We thank M.\ Lublinsky (Ben-Gurion University of the Negev) and the
organizers of the workshop ``Physics of Saturation -- precision and
quasicollectivity'' for their support and hospitality in May 2022; we
are also grateful to the participants of the workshop for stimulating
discussions which initiated this work.

\appendix
\section{Shannon entropy of a probability density function for
  a continuous degree of freedom}
\label{sec:app_Shannon}

In this appendix we review the definition of the entropy associated
with a classical probability density function over a continuous degree
of freedom.  We discuss the extension to a quantum mechanical density
matrix at the end.\\

First, recall the expression for the classical Shannon entropy
for a set of {\em discrete} outcomes of a random draw, with probabilities
$P_i$:
\be
H(\{P_i\}) = - \sum_i P_i \log P_i~.
\ee
If the set of possible outcomes is {\em continuous}, e.g.\ $x\in
\mathbb{R}^+$, their distribution is given by a normalized, integrable
(including possibly the $\delta$-function measure) probability density
function $p(x)>0$,
\be
\int\dd x\, p(x) = 1~.
\ee
Note that if $x$ is dimensional then $\mathrm{dim}(p)=
[\mathrm{dim}(x)]^{-1}$.  Also, that the integration measure does not
involve {\em any $x$-dependent Jacobians}, all of which must be
absorbed into $p(x)$ for it to be a valid probability density with
respect to the integration measure $\dd x$.

To apply Shannon's formula here, we first discretize the continuous set
of outcomes by introducing (equal size)
bins $\Delta x>0$. An outcome $x$ falls
into bin $i = \lfloor{x/\Delta x}\rfloor$. The probability $P_i$ for
an event in bin $i$, is
\be
P_i = \int\limits_{i\Delta x}^{(i+1)\Delta x} \dd x\, p(x)
\equiv p_i\, \Delta x~,
~~~~~~~~~~~(i\in \mathbb{N}_0)~.
\ee
In the last step we defined the binned density $p_i$ as the average
of the probability density $p(x)$ over bin $i$.

We now have
\be \label{eq:H[p]_sum}
H[p] = - \sum_i \int\limits_{i\Delta x}^{(i+1)\Delta x} \dd x\, p(x)\,
\log(p_i\, \Delta x)~.
\ee
If $p(x)$ is a continuous function then
\be
H[p] = - \int \dd x\, p(x)\,
\log(p(x)\, \Delta x)~,
\ee
and the entropy does not have a finite $\Delta x\to0$ limit.
(Even so, the relative entropy for two such probability densities
does converge.)
If, on the other hand, $p(x)$ is given by a sum of Dirac $\delta$
functions then eq.~(\ref{eq:H[p]_sum}) does converge since this
basically recovers the case of discrete outcomes.\\

Along similar lines, let $\rho_{xx'}$ denote a density matrix
describing a continuous degree of freedom. By postulate, its trace
is normalized,
\be
\tr\,  \rho = \int\dd x \, \rho_{xx} =
\int\frac{\dd x}{\Delta x} \,\, (\Delta x)\, \rho_{xx} =
1~.
\ee
Once again we stress that the trace measure must be $\dd x$, with any
$x$-dependent Jacobians absorbed into $\rho$. The von~Neumann entropy
$S$ is given by the Shannon entropy of the vector of dimensionless
eigenvalues of $(\Delta x)\, \rho$.  Upon binning the eigenvalue
distribution, it is given by
\be
S = - \sum_i \int\limits_{i\Delta \lambda}^{(i+1)\Delta \lambda}
\dd \lambda\, p(\lambda)\,
\log(p_i\, \Delta \lambda)~.
\ee
%

\section{Calculating traces of powers of $\rho$}

Gearing up for the calculation of the von Neumann entropy, we write
expressions for the trace of powers of the density matrix, $\tr\,
(\rho_{\overline A})^N$ (which, if desired can be used to determine
the Renyi entropy).

Since our density matrix is block diagonal in the particle number
basis, the different blocks don't talk to each other and can be
considered separately.

In the zero particle subspace $\rho_0$ is a number and thus
\begin{equation}\label{0}
\tr\, \rho_0^N=\rho_0^N ~.
\end{equation}

It is easy to see that in the three particle subspace
we simply have
\be \label{3}
\tr\,\rho_3^N = (\tr\,\rho_3)^N =
\left[\int[\dd x_i]\,[\dd^2r_i]\,
  \Theta_{\overline A}(\vec r_1)\,\Theta_{\overline A}(\vec
r_2)\,\Theta_{\overline A}(\vec r_3)\,\, |\Psi(x_i,\vec r_i)|^2
  \right]^N~.
\ee
Note that in both \eqref{0} and \eqref{3} the lattice spacing $a$ does
not appear.\\

Consider now the single particle subspace:
\bea
(\rho_1^{\, 2})_{\alpha \alpha} &=& (\rho_1)_{\alpha \alpha} \,
(\rho_1)_{\alpha \alpha} \\
&=& 3\Delta x \,a^2 \int
\frac{\dd x_1\dd x_2}{8 x_1 x_2 x_3}\,\delta(1-x_1-x_2-x_3)\,
\int\dd^2 r_1\, \dd^2r_2\, \delta(x_1\vec r_1+x_2\vec r_2+x_3\vec r_3)
\,\Theta_A(\vec r_1)\,\Theta_A(\vec r_2)\,
|\Psi(x_1,\vec r_1; x_2,\vec r_2; x_3,\vec r_3)|^2 \nn\\
& & 3\Delta x \, a^2\int
\frac{\dd y_1\dd y_2}{8 y_1 y_2 x_3}\,\delta(1-y_1-y_2-x_3)\,
\int\dd^2 s_1\, \dd^2s_2\, \delta(y_1\vec s_1+y_2\vec s_2+x_3\vec r_3)
\,\Theta_A(\vec s_1)\,\Theta_A(\vec s_2)\,
|\Psi(y_1,\vec s_1; y_2,\vec s_2; x_3,\vec r_3)|^2~.\nonumber
\eea
Taking the trace,
\bea
\tr\, \rho_1^{\, 2} &=&
\int\frac{\dd x_3}{\Delta x} \int\frac{\dd^2r_3}{a^2}\,
\Theta_{\overline A}(\vec r_3)\nn\\
& &
   3\Delta x \, a^2 \int
   \frac{\dd x_1\dd x_2}{8 x_1 x_2 x_3}\,\delta(1-x_1-x_2-x_3)\,
   \int\dd^2 r_1\, \dd^2r_2\, \delta(x_1\vec r_1+x_2\vec r_2+x_3\vec r_3)
\,\Theta_A(\vec r_1)\,\Theta_A(\vec r_2)\,
|\Psi(x_1,\vec r_1; x_2,\vec r_2; x_3,\vec r_3)|^2 \nn\\
& & 3\Delta x \, a^2\int
\frac{\dd y_1\dd y_2}{8 y_1 y_2 x_3}\,\delta(1-y_1-y_2-x_3)\,
\int\dd^2 s_1\, \dd^2s_2\, \delta(y_1\vec s_1+y_2\vec s_2+x_3\vec r_3)
\,\Theta_A(\vec s_1)\,\Theta_A(\vec s_2)\,
|\Psi(y_1,\vec s_1; y_2,\vec s_2; x_3,\vec r_3)|^2 \nonumber\\
&=&
\int\frac{\dd x_3}{\Delta x}\int\frac{\dd^2 r_3}{a^2}
\,\Theta_{\overline A}(\vec r_3)\,
\left[3 \Delta x\, a^2\,
  \int[\dd y_i]\,\delta(y_3-x_3)\int [\dd^2 s_i]\, \delta(\vec s_3 - \vec r_3)
\,\Theta_A(\vec s_1)\,\Theta_A(\vec s_2)\,
|\Psi(y_i,\vec s_i)|^2
  \right]^2
~.
\label{eq:rho1^2}
\eea
For an arbitrary $N$ we obtain
\bea
\tr\, \rho_1^{\, N} &=& \int\frac{\dd x_3}{\Delta x} \int\frac{\dd^2r_3}{a^2}\,
\Theta_{\overline A}(\vec r_3)\,
\left[3\Delta x\, a^2 \int[\dd y_i] \,\delta(y_3-x_3)
  \int[\dd^2 s_i]\, \delta(\vec s_3 - \vec r_3)
\,\Theta_A(\vec s_1)\,\Theta_A(\vec s_2)\,
|\Psi(x_i,\vec s_i)|^2\right]^{N}\!.
\label{eq:rho1^N}
\eea
As noted in the main text of this paper, the lattice spacing does not cancel in this
expression, and formally this expression vanishes, for $N>1$, in the
``continuum limit" $\Delta x, a\rightarrow 0$.
\\

Now consider traces of powers of $\rho_2$. We have
\bea
(\rho_2^{\, 2})_{\alpha \alpha'} &=& 9 \left(\Delta x\, a^2\right)^3\,
\int[\dd y_i]\,[\dd^2 s_i] \, \Theta_{\overline A}(\vec s_1)\,
\Theta_{\overline A}(\vec s_2)\,
\delta(x_3-y_3)\,
\delta(\vec s_3-\vec r_3)\,\,
\left|\Psi(y_i,\vec s_i)\right|^2  \nn\\
& &~~~~~~~~ \times
\frac{\Psi(x_i,\vec r_i)}{x_3\sqrt{x_1 x_2 x_3}}\,\,
\frac{\Psi^*(x_i',\vec r_i\!')}{x_3\sqrt{x_1' x_2' x_3}}~.
\eea
In this expression $\vec r_3=-(x_1\vec r_1+x_2\vec
r_2)/x_3 = \vec r_3\!'=-(x_1'\vec r_1\!'+x_2'\vec r_2\!')/x_3'$, while $x_3=
1-x_1-x_2 = x_3'=1-x_1'-x_2'$. Recall, also, that here the indices
$\alpha=\{x_1,\vec r_1, x_2, \vec r_2\}$, $\alpha'=\{x_1',\vec r_1\!',
x_2', \vec r_2\!'\}$, are defined over the domain $\vec r_1, \vec r_2,
\vec r_1\!', \vec r_2\!' \in \overline A$, $\vec r_3\in A$.

The trace, defined with the measure $\dd x_1\,\dd x_2\, \dd^2r_1\,
\dd^2 r_2 \, \Theta(x_3)\, \Theta_A(\vec r_3)\,\Theta_{\overline A}(\vec
r_1)\,\Theta_{\overline A}(\vec r_2)\, / (a^2\Delta x)^2$, is
\bea
\tr\, \rho_2^{\, 2} &=&
3 \int [\dd x_i]\,[\dd^2r_i]\, \Theta_A(\vec
r_3)\,\Theta_{\overline A}(\vec r_1)\,\Theta_{\overline A}(\vec r_2)\,
|\Psi(x_i,\vec r_i)|^2 \nn\\
& & \times 3\Delta x\, a^2
\int[\dd y_i]\,[\dd^2 s_i] \, \Theta_{\overline A}(\vec s_1)\,
\Theta_{\overline A}(\vec s_2)\, \delta(x_3-y_3)\,\delta(\vec s_3-\vec r_3)\,
|\Psi(y_i,\vec s_i)|^2~.
\eea
Note that this is actually identical to eq.~(\ref{eq:rho1^2}) for
$\tr\, \rho_1^{\, 2}$ with $A\leftrightarrow\overline A$, as it should
be.

For general power $N$,
\bea
\tr\, \rho_2^{\, N} &=&
\int\frac{\dd x_3}{\Delta x} \int\frac{\dd^2r_3}{a^2}\,
\Theta_{A}(\vec r_3)\,
\left[3\Delta x\, a^2 \int[\dd y_i]\,[\dd^2 s_i]\,
  \delta(x_3-y_3)\, \delta(\vec s_3 - \vec r_3)
\,\Theta_{\overline A}(\vec s_1)\,\Theta_{\overline A}(\vec s_2)\,
|\Psi(y_i,\vec s_i)|^2\right]^{N}~.
\label{eq:rho2^N}
\eea
Again we see that the lattice spacing does not disappear in this
expression and formally leads to its vanishing for $a\rightarrow 0$
and $N>1$.

\section {Checking traces}  \label{sec:app-traces}

Here we
check the normalization of the density matrix \eqref{eq:rhoNLO-traced}.

First let us take the trace of  $\rho^{qqq}$, which amounts to setting
$x_i'=x_i$, $\vec r_i\!'=\vec r_i$, and integrating over $[\dd x_i]$ and
$[\dd^2 r_i]$. The first line gives
\be
\int[\dd x_i]\,[\dd^2 r_i]\, \left(1-3C^{\mathrm{reg}}_q(x_1)\right)\,\,
\left| \Psi_\mathrm{qqq}(x_i, \vec r_i) \right|^2 =
1 - 3 \int[\dd x_i]\,[\dd^2 r_i]\,\, C^{\mathrm{reg}}_q(x_1)\,\,
\left| \Psi_\mathrm{qqq}(x_i, \vec r_i) \right|^2
~,  \label{eq:tr_rho-qqq_diag}
\ee
where we used the normalization condition~(\ref{eq:norm_tr-Abar})
for the 3-quark wave function. From the rest of eq.~(\ref{eq:rho_qqq_virt})
we get
\bea
& &
2g^2 C_F\,\int[\dd x_i]\,[\dd^2 r_i]\,
\int_{\Delta x} \frac{\dd x_g}{2x_g}\frac{\dd^2 k_g}{(2\pi)^3}\,
\frac{k_g^2}{(k_g^2+\Delta^2)^2}\,\,
\left|\Psi_\mathrm{qqq}\left(x_i,\vec r_i\right)\right|^2\,\,
\notag \\ & & \times
\left[
e^{-i\vec k_g\cdot(\vec r_1-\vec r_2)}\,
+
e^{-i\vec k_g\cdot(\vec r_1-\vec r_3)}\,
+
e^{-i\vec k_g\cdot(\vec r_2-\vec r_3)}\,
+ \mathrm{c.c.}\right]~.
\label{eq:tr_rho-qqq_offdiag}
\eea
\\

Now let us take the trace of the matrix
\eqref{eq:rho_qqqg_aa'-line2}. To do this set $x_i'=x_i$, $\vec
r_i\!'=\vec r_i$, and integrate over all degrees of freedom, including
the momentum fraction of the gluon with the measure $\dd x_g/2x_g$,
and its transverse position with the measure $2\pi\, \dd^2 r_g$. This
is done by performing the following steps: i) extract the integrations
over $\dd x_g/2x_g$, $\dd^2 k_g/(2\pi)^3$, and $\dd^2 k_g'/(2\pi)^3$
from $[\dd x_i]$, $[\dd^2 k_i]$, and $[\dd^2 k_i']$; ii) perform the
integration over $\dd^2 r_g$ which produces a $(2\pi)^2\, \delta(\vec
k_g-\vec k_g')$; iii) shift the quark momenta (as needed) by $-\vec
k_g$ so that the arguments of the $\Psi_\mathrm{qqq}$ functions no
longer involve $\vec k_g$; note that this also changes $\delta(\vec
k_1+\vec k_2+\vec k_3+\vec k_g) \to \delta(\vec k_1+\vec k_2+\vec
k_3)$, and similar for the primed momenta.  The first three terms of
eq.~(\ref{eq:rho_qqqg_aa'-line1}) then give (taking $\Delta^2\to0$
where possible)
\bea
{3\cdot 4g^2C_F} \, \int[\dd x_i]\, [\dd^2 r_i]\,\,
\left| \Psi_\mathrm{qqq}(x_i,\vec r_i)\right|^2\,
\int_{\Delta x}^{x_1}
\frac{\dd x_g}{2x_g}\int\frac{\dd^2 k_g}{(2\pi)^3} \frac{1}{k_g^2+\Delta^2}
~.
\eea
This cancels against the ${\cal O}(g^2)$ correction in
eq.~(\ref{eq:tr_rho-qqq_diag}), after regularization of the UV
divergence, $(k_g^2+\Delta^2)^{-1} \to (k_g^2+\Delta^2)^{-1} -
(k_g^2+\Lambda^2)^{-1}$.

The remaining terms of~(\ref{eq:rho_qqqg_aa'-line2}) give
\bea
& &
-{2g^2 C_F}\,\int[\dd x_i]\,[\dd^2 r_i]\,\,
\left|\Psi_\mathrm{qqq}\left(x_i,\vec r_i\right)\right|^2\,
\int_{\Delta x} \frac{\dd x_g}{2x_g}\frac{\dd^2 k_g}{(2\pi)^3}\,
\frac{k_g^2}{(k_g^2+\Delta^2)^2}\,
\notag \\ & & \times
\left[
e^{-i\vec k_g\cdot(\vec r_1-\vec r_2)}\,
+
e^{-i\vec k_g\cdot(\vec r_1-\vec r_3)}\,
+
e^{-i\vec k_g\cdot(\vec r_2-\vec r_3)}\,
+ \mathrm{c.c.}\right]~.
\eea
This cancels against~(\ref{eq:tr_rho-qqq_offdiag}).  The cancellations
of the perturbative corrections in the trace ensure that it remains =1,
and independent of the coupling $g^2$ and the
IR (collinear and soft) and UV cutoffs.

\bibliography{refs}

\end{document}